\documentclass[sn-mathphys,Numbered, iicol]{sn-jnl}

\usepackage{graphicx}%
\usepackage{multirow}%
\usepackage{amsmath,amssymb,amsfonts}%
\usepackage{amsthm}%
\usepackage{mathrsfs}%
\usepackage[title]{appendix}%
\usepackage{xcolor}%
\usepackage{textcomp}%
\usepackage{manyfoot}%
\usepackage{booktabs}%
\usepackage{algorithm}%
\usepackage{algorithmicx}%
\usepackage{algpseudocode}%
\usepackage{listings}%

\usepackage{subcaption}
\usepackage{multicol}

\usepackage{makecell}

\lstdefinestyle{malwarelistingstyle}{
  backgroundcolor=\color{blue!8},         
  basicstyle=\ttfamily\scriptsize,        
  breaklines=true,
  breakatwhitespace=false,
  columns=fullflexible,
  frame=single,
  showstringspaces=false,
  tabsize=4,
  keepspaces=true,
  linewidth=\textwidth
}

\usepackage{xcolor}%

\begin{document}

\title[Article Title]{RawMal-TF: Raw Malware Dataset Labeled by Type and Family}


\author[1]{\fnm{David} \sur{B\'{a}lik}}\email{balikda1@fit.cvut.cz}

\author*[1]{\fnm{Martin} \sur{Jure\v{c}ek}}\email{martin.jurecek@fit.cvut.cz}

\author[2]{\fnm{Mark} \sur{Stamp}}\email{mark.stamp@sjsu.edu}

\affil*[1]{\orgdiv{Faculty of Information Technology}, \orgname{Czech Technical University in Prague},\\ \orgaddress{\city{Prague}, \country{Czechia}}}
\affil[2]{\orgdiv{Department of Computer Science}, \orgname{San Jose State University}, \orgaddress{\city{San Jose}, \state{California}, \country{USA}}}



\abstract{This work addresses the challenge of malware classification using machine learning by developing a novel dataset labeled at both the malware type and family levels. Raw binaries were collected from sources such as VirusShare, VX Underground, and MalwareBazaar, and subsequently labeled with family information parsed from binary names and type-level labels integrated from ClarAVy. The dataset includes 14 malware types and 17 malware families, and was processed using a unified feature extraction pipeline based on static analysis, particularly extracting features from Portable Executable headers, to support advanced classification tasks. The evaluation was focused on three key classification tasks. In the binary classification of malware versus benign samples, Random Forest and XGBoost achieved high accuracy on the full datasets, reaching 98.5\% for type-based detection and 98.98\% for family-based detection. When using truncated datasets of 1,000 samples to assess performance under limited data conditions, both models still performed strongly, achieving 97.6\% for type-based detection and 98.66\% for family-based detection. For interclass classification, which distinguishes between malware types or families, the models reached up to 97.5\% accuracy on type-level tasks and up to 93.7\% on family-level tasks. In the multiclass classification setting, which assigns samples to the correct type or family, SVM achieved 81.1\% accuracy on type labels, while Random Forest and XGBoost reached approximately 73.4\% on family labels. The results highlight practical trade-offs between accuracy and computational cost, and demonstrate that labeling at both the type and family levels enables more fine-grained and insightful malware classification. The work establishes a robust foundation for future research on advanced malware detection and classification.}




\maketitle

\section{Introduction}\label{sec1}

Malware remains one of the most pervasive and evolving threats in today's digital world. From ransomware attacks targeting critical infrastructure to sophisticated nation-state campaigns, the landscape of malicious software continues to grow in both scale and complexity. As the volume of new malware samples rises daily, the need for automated, data-driven detection approaches has become increasingly urgent. Machine-learning (ML) models have shown remarkable promise in this domain \cite{ucci2019survey}, offering the ability to automatically learn patterns from data and generalize to unseen threats.

However, the success of ML-based malware detection systems depends critically on the quality and completeness of the data used for training. The more precise and representative the dataset, the better the models that can be built \cite{dved2025selecting}. Many existing malware datasets, such as the widely used EMBER dataset \cite{2018arXiv180404637A}, provide only precomputed feature vectors extracted from binaries, typically summarizing static properties like Portable Executable (PE) headers, imported functions, or byte histograms. While such feature datasets are useful for reproducibility and benchmarking, they come with significant limitations.

This work addresses these limitations by presenting a comprehensive malware dataset that, unlike many existing resources, includes the raw binary files alongside their extracted features and family and type labels. Access to binary samples is critical because it enables researchers not only to extract custom or updated feature representations, but also to investigate advanced use cases such as adversarial attacks, reverse engineering, and dynamic analysis; capabilities that are impossible to explore with precomputed feature vectors alone. Furthermore, the availability of raw binaries provides a critical advantage: it allows researchers to apply new or improved feature extraction techniques in the future, adapt to evolving analysis needs, and explore alternative machine learning strategies. This flexibility ensures that the dataset remains useful even as tools and methods advance, avoiding the limitations of datasets that only provide precomputed features and lock researchers into the choices made during an earlier preprocessing stage.

The dataset used in this work was assembled by combining multiple publicly available sources. For malware types, precomputed labels from ClarAVy \cite{joyce2025claravy} were leveraged; these labels were available for binaries hosted on VirusShare \cite{virusshare}, providing ready-made annotations derived from large-scale antivirus scan data. For malware families, samples were collected from VX Underground \cite{vxunderground}, which aggregates malware batches originating from sources like MalwareBazaar \cite{malwarebazaar}. These binaries often included the family name directly in their filenames, which was systematically parsed to assign the corresponding family labels. Malware samples labeled by type were collected from the MalwareBazaar collection on VX-Underground for the period from January 2023 to March 2025. Malware samples labeled by family were obtained from malware packages uploaded to VirusShare between October 10, 2023, and March 26, 2025. Note that each sample includes the TimeDateStamp field from the COFF File Header, which indicates when the file was created; however, this field can be easily modified. 

This combined approach made it possible to build a diverse and well-labeled dataset without relying on expensive or restricted services like the VirusTotal API \cite{virustotal}, ensuring both scalability and consistency for the experiments. The dataset used in this study, which includes malware binaries and the corresponding feature vectors labeled at both the malware type and family levels, is available at the following link: \url{https://github.com/CS-and-AI/RawMal-TF}.

Beyond dataset construction, this work carries out a comprehensive series of experiments to evaluate the effectiveness of machine-learning models for malware classification. Multiple tasks were addressed, including binary detection of malware versus benign samples, interclass classification between different malware types or families, and multiclass categorization where models predict the precise malware category. Both full-scale and reduced datasets were used to systematically investigate how data volume, balance, and label granularity affect model performance. The experimental setup was designed to reflect realistic challenges faced by malware detection systems and to generate practical insights into the behavior of widely used machine-learning algorithms under different conditions.\\

The rest of this paper is organized as follows. Section 1 introduces the necessary background, including an overview of malware, its main types and families, labeling techniques using tools such as AVClass and ClarAVy, a summary of relevant datasets including EMBER, and an overview of the machine learning algorithms used. Section 2 describes the dataset collection and preparation process, covering the acquisition of binary samples, the application of labels, the initial and final dataset construction approaches, and the creation of type-based and family-based datasets. Section 3 presents the experimental work, including the feature extraction pipeline, the design of binary and multiclass classification tasks, and the evaluation of machine-learning models. We analyze the results and compare the performance of different models across various classification tasks. In the Conclusion, we summarize the main contributions of the work, discuss limitations, and propose directions for future research.

\section{Background}\label{sec2}

\subsection{Malware Overview}
Malware, short for malicious software, is a general term for software that is intentionally created to perform unauthorized or harmful actions on a computing system. It includes various categories such as viruses, worms, trojans, ransomware, spyware, and others \cite{nist7621r1}. The typical objectives of malware are data theft, system disruption, unauthorized access, or control over compromised machines \cite{avastMalware}.


In practical terms, malware is usually defined not just by its static characteristics, but by its behavior and context. This behavioral perspective is essential in modern classification techniques, particularly those based on machine learning, which rely on patterns observed during execution or extracted from code features in combination with high-quality labeled data \cite{huang2024data}.

\subsubsection{Malware Types}
Malware can be classified by its high level purpose or by observed functionality. In practice, however, many modern threats combine several behaviors, making a single functional label difficult to assign. As shown in \cite{lanier2025worm, malwarebytesDownloader}, a given sample may perform network scanning and vulnerability exploitation, activities typical for worms, while also downloading and installing additional components, behaviors associated with downloaders. As a result, the boundaries between types such as worms and downloaders often overlap, complicating taxonomy design and detection strategies.

Malware types, as described in \cite{malwarebytesDownloader, nist80083r1, nist8374draft, nistSpyware, kasperskyAdware}, categorize samples by their primary behavior or mechanism of action. Examples include:

\begin{itemize}
  \item \emph{Viruses}, which attach to legitimate executables and propagate when they are executed.
  \item \emph{Worms}, which self replicate across networks without user intervention.
  \item \emph{Trojans}, which masquerade as benign software to trick users into execution.
  \item \emph{Ransomware}, which encrypts or locks data to extort payment.
  \item \emph{Spyware}, which stealthily collects user or system information.
  \item \emph{Adware}, which displays unwanted advertisements to generate revenue.
  \item \emph{Droppers and downloaders}, which serve as lightweight stagers that fetch and install additional payloads.
\end{itemize}

Type labels focus on high level behavior such as data theft, disruption, or credential theft, guiding general detection policies but not revealing implementation lineage or code reuse.

\subsubsection{Malware Families}
Malware families group samples sharing a common code base, development origin, or toolkit \cite{joyce2021motif}. Family members evolve through versions yet retain identifiable artifacts such as unique encryption routines or network beacons. For instance, the \emph{Emotet} family began as a banking trojan with specific command and control protocols  and later incorporated self propagation modules \cite{trendmicro2020emotet}, while \emph{TrickBot} maintains a modular architecture for credential theft and lateral movement \cite{constantinTrickbot}. Family labels emphasize lineage and attribution, critical for tracking campaigns and crafting targeted defenses.

Common malware families, as described in \cite{microsoftAllaple, huntioEmotet, esetVirlock, microsoftVirut, huntioNjrat, fsecureZbot}, exhibit varied propagation and payload mechanisms. Examples include:
\begin{itemize}
  \item \emph{Allaple}: polymorphic network worm that spreads to other systems and can perform denial of service attacks
  \item \emph{Emotet}: modular Trojan delivered via phishing that propagates laterally by cracking credentials and fetching additional malware 
\item \emph{Virlock}: polymorphic ransomware-virus whose body changes with each file and execution, employing multiple layers of encryption to lock the screen and infect files

  \item \emph{Virut}: virus with backdoor features communicating via IRC channels
 \item \emph{njRAT} (Bladabindi): remote access Trojan, offering modular control features and extensive tutorials for easy deployment
  \item \emph{Zbot} (Zeus): banking trojan that captures web credentials 
\end{itemize}

\subsection{Labeling Software}
In malware research, accurate labeling of samples is essential for training and evaluating detection systems. Labeling software refers to tools designed to automatically assign meaningful tags such as family names or behavioral categories to malware binaries based on various sources of metadata. These tools help process large datasets by extracting the most likely classification from a~collection of noisy or inconsistent labels provided by antivirus engines.

\subsubsection{AVClass}
AVClass \cite{sebastian2016avclass} is a Python-based tool designed for automatic malware labeling by analyzing antivirus detection labels. By providing AV labels for a large set of malware samples (for example, VirusTotal JSON reports), AVClass identifies the most probable family name for each sample and can extract additional tags describing malware class, behaviors, and file properties.

One of the main strengths of AVClass is its ability to process large datasets efficiently without manual intervention. It is vendor-agnostic, meaning it can utilize labels from any available AV engines, even when these vary between samples. Furthermore, it supports multiple platforms, including Windows and Android, and does not require access to the executable files themselves-only the detection labels are needed, which can often be obtained using a sample's hash.

%
%

\subsubsection{ClarAVy}
ClarAVy \cite{joyce2025claravy} is a Python-based command-line tool designed to analyze antivirus detection reports and assign detailed labels to malware samples. It extends beyond simple family labeling by providing additional categorizations such as behavioral tags (e.g., ransomware, downloader), file properties (e.g., ELF, PDF, Java), exploited vulnerabilities (e.g., CVE-2017-0144), packers (e.g., UPX, Themida), and threat actor associations (e.g., Lazarus Group, APT1). ClarAVy outputs a confidence score for each family prediction, providing a measure of certainty in its classification.

%

ClarAVy differs from AVClass primarily through its advanced parsing techniques. It recognizes over 900 antivirus label formats from 103 supported AV engines, using pattern matching and context-aware token classification to infer label meanings. It achieves high parsing coverage (99.5\%) across large datasets, and applies a Variational Bayesian approach to infer the most probable family classification.

%

\subsection{EMBER}

EMBER (Elastic Malware Benchmark for Empowering Researchers) \cite{2018arXiv180404637A} is a benchmark dataset and feature extraction framework designed for static analysis of Windows Portable Executable (PE) files. It provides not only labeled datasets but also a standardized Python-based pipeline for extracting and vectorizing features from arbitrary PE files.

\subsubsection{Feature Extraction}
The EMBER dataset and its tools are built on top of the LIEF (Library to Instrument Executable Formats) project \cite{lief}, which allows detailed parsing of PE files. In the context of this work, the official EMBER extraction pipeline was utilized to generate raw feature representations from custom-collected binaries. 

%


The extracted features include a wide range of static characteristics of PE files, and only selected examples from each feature category are presented here to illustrate the general structure. A complete description of all features is available in \cite{2018arXiv180404637A}.

\begin{itemize}
\item \textbf{Header features:} Number of sections, timestamps, and import/export table metrics.
\item \textbf{Byte-level histograms:} Including both flat byte distributions and entropy-based 2D histograms.
\item \textbf{Section properties:} Names, sizes, entropy values, and permissions.
\item \textbf{String analysis:} Frequencies of printable strings, URLs, file paths, and registry keys.
\item \textbf{Imported functions:} Names and frequencies of Windows API calls.
\end{itemize}

Researchers can also extend or re-vectorize these features to suit specific experimental needs.

\subsection{Classifiers}
In this work, five widely used machine learning classifiers were applied to the problem of malware detection and classification, including K-Nearest Neighbors (KNN), Multi-Layer Perceptron (MLP), Support Vector Machine (SVM), Extreme Gradient Boosting (XGBoost), and Random Forest.

\subsubsection{K-Nearest Neighbors (KNN)}
K-Nearest Neighbors (KNN) is a simple instance-based learning algorithm that stores representations of training points and, when classifying a new sample, identifies the $k$ nearest training neighbors in the feature space. The class is assigned based on majority voting among these neighbors, i.e., the new object is classified into the class most represented among them. KNN is a lazy learning approach with no model parameters to estimate, making it easy to implement; however, it can become computationally expensive at prediction time, especially on large datasets.


\subsubsection{Multi-Layer Perceptron (MLP)}
The Multi-Layer Perceptron (MLP) is a feedforward neural network consisting of an input layer, one or more hidden layers of neurons, and an output layer. Each hidden unit performs a weighted sum of its inputs, adds a bias term, and applies a nonlinear activation function. Thanks to these nonlinear layers, MLPs can approximate complex decision functions and model intricate feature relationships. Training is done iteratively using the backpropagation algorithm to adjust the weights and minimize prediction error on the training data. MLPs are thus universal approximators capable of learning both linear and highly nonlinear dependencies.


\subsubsection{Support Vector Machine (SVM)}
The Support Vector Machine (SVM) is a classification algorithm that seeks the optimal decision hyperplane that maximizes the margin (separation) between data classes. The SVM model relies on a subset of the training samples, the support vectors, which are the points closest to the decision boundary and define its position. As a result, SVMs are memory-efficient and effective even in high-dimensional spaces. Using kernel functions (kernel trick), SVMs can be extended to handle nonlinearly separable tasks: the input feature space is mapped into a higher-dimensional space where linear separation becomes possible. By selecting different kernels (linear, polynomial, Gaussian RBF, etc.), SVMs become versatile classifiers for various data types.


\subsubsection{Extreme Gradient Boosting (XGBoost)}
XGBoost is a state-of-the-art gradient boosting algorithm that builds an ensemble of weak learners, typically decision trees, by sequentially optimizing a differentiable loss function using gradient descent. The model starts with a simple base learner and incrementally adds new trees that focus on correcting the residual errors of the previous iteration. XGBoost generalizes boosting frameworks like AdaBoost by supporting arbitrary differentiable loss functions and incorporates regularization (L1 and L2) to prevent overfitting. It also introduces stochastic elements, such as subsampling of training instances and features at each iteration, which help reduce overfitting and improve generalization, similar to techniques used in bagging.

\subsubsection{Random Forest}
Random Forest is a popular ensemble algorithm that combines many decision trees trained using bootstrap aggregation (bagging). Each tree in the forest is trained on a different random subset of the training data (bootstrap samples), and during tree construction, only a random subset of features is considered at each split. This double randomness (in data and features) ensures that the resulting trees are less correlated with each other. By combining many such trees, typically through voting or averaging, the variance of the model is significantly reduced compared to individual trees, which tend to overfit. Random Forests usually achieve high accuracy without requiring extensive hyperparameter tuning and provide feature importance measures, making the models easier to interpret. The main downside is slightly higher inference time (due to traversing many trees), but this is often outweighed by the model's improved accuracy.


\section{Dataset Collection and Preparation}\label{Dataset Collection and Preparation}

There are very few publicly available datasets that combine actual malware binaries with consistent, large-scale labeling at both the type and family level. Datasets with this level of detail and structure are rare, making such resources especially valuable for research in malware analysis and classification. This section describes the process of constructing such a dataset using real-world malware samples collected from public repositories. In our dataset each sample is uniquely identified and categorized according to its general behavior (e.g.,~worm, trojan, adware) as well as its specific malware family (e.g., Zbot, Allaple, Emotet).

Several approaches were explored during this process, including methods that relied on the aggregation of antivirus scan results. In particular, an initial attempt to retrieve detection reports from VirusTotal and process them using labeling tools like AVClass proved to be impractical when using the free public API, due to strict API rate limits. These constraints made it infeasible to label a large number of samples within a reasonable timeframe. As a result, alternative strategies had to be considered, leading to the adoption of a more scalable and consistent approach based on ClarAVy labels.

\subsection{Obtaining Data}
To construct a dataset suitable for machine learning-based malware classification, it is necessary to obtain access to real-world malware samples. In this work, three primary sources were used for this purpose: \textbf{VirusShare, Vx Underground  and MalwareBazaar}. Each of these repositories played a~specific role in building a dataset that supports classification by both malware type and family.

\begin{description}
\item[\hypertarget{virusshare}{VirusShare}]\cite{virusshare} served as the primary source of raw malware binaries for samples that were later categorized by their \textbf{behavioral type}. The repository offers a large collection of malware samples organized into sequentially numbered \texttt{.zip} archives, which are distributed via torrent. As of writing, 487 such archives are available for download, totaling approximately 16,791 GB of data. However, VirusShare does not provide any labels or metadata for the contained binaries.



\item[MalwareBazaar]\cite{malwarebazaar} was initially considered for the \textbf{family-level} classification of malware. This platform offers access to recent malware samples along with structured metadata, including family names, antivirus labels, classification tags, and YARA rule matches. MalwareBazaar offers a structured API that allows to query and download samples along with their metadata. However, during the period when data collection for this work was performed, the API was not functioning reliably, making it infeasible to obtain a large number of samples directly.

\item[\hypertarget{vxunderground}{VX Underground}]\cite{vxunderground} was used as an alternative source for obtaining malware binaries, particularly in cases where direct access through other platforms was not feasible. 
Although Vx Underground does not provide structured metadata or a complete mirror of MalwareBazaar, it hosts a significant number of curated malware datasets, including partial exports from MalwareBazaar. One such collection, titled \textit{Bazaar Collection / Downloadable Releases}, contains month-wise compressed archives (e.g., Bazaar.2025.02.7z) with malware samples categorized by date. 
While this dataset is not exhaustive, it served as a practical and scalable way to retrieve large amounts of malware samples for family-level classification when direct use of the MalwareBazaar API was not possible.

\end{description}

\subsection{Initial Approach}
\label{sec:initial-approach}

 The initial strategy considered for constructing the labeled dataset involved retrieving antivirus scan reports from VirusTotal and using them to infer malware type or family label. As of the time of writing, VirusTotal integrates outputs from 73 antivirus engines, offering a comprehensive view of how different vendors classify given sample. In this strategy, the input samples were selected from the VirusShare, which served as the source of raw, unlabeled malware binaries intended for subsequent processing through the VirusTotal API.

 The scan reports were stored in \texttt{.jsonl} (JSON Lines) format, which is supported by AVClass as a third party labeling software. Each JSON object contains three main components:

\begin{itemize}
    \item \texttt{data} - the root object containing sample-specific information;
    \item \texttt{attributes} - a nested object holding detailed metadata such as detection results, file properties, submission history, and popularity indicators;
    \item \texttt{last\_analysis\_results} - a dictionary mapping antivirus engine names to their individual scan verdicts, including classification labels, engine metadata, detection category (e.g., malicious, undetected), and other attributes.
\end{itemize}

The \texttt{last\_analysis\_results} component is the primary input for AVClass. It contains raw detection names as returned by antivirus engines, which often differ significantly across vendors in both syntax and terminology. For example, a single sample may be labeled as \texttt{Trojan.Zbot, Win32/Zbot.B}, or \texttt{Malware.Generic}, depending on the engine. Due to this inconsistency, the detection names cannot be directly used for classification or evaluation purposes. AVClass is used to resolve this issue by normalizing and aggregating the labels across vendors, enabling the assignment of consistent and standardized malware family names. In addition to family-level classification, AVClass can also be used to infer general malware types.

Despite conceptual viability of this labeling approach, it proved infeasible for large-scale use due to strict limitations imposed by the VirusTotal public API. The free tier allows a maximum of 4 requests per minute and 500 requests per day \cite{virustotal}, which makes the processing of hundreds of thousands  of samples impractical within any reasonable timeframe.


The second major limitation of this approach is the absence of any prior knowledge regarding the distribution of malware families or types within the VirusShare archives. Since these archives served as the source of samples in this labeling strategy, the absence of any labeling within the VirusShare binaries made it impossible to control for class balance or diversity prior to submission. Each archive contains tens of thousands of unlabeled binaries, and family or type information becomes available only after a sample is processed through VirusTotal and AVClass. This introduces a substantial risk that the resulting dataset will be biased toward a few dominant malware types or families, while other classes may be underrepresented or entirely absent.

As a result, the use of VirusTotal scan data for large-scale labeling had to be abandoned. An alternative method based on pre-labeled datasets was selected for further dataset construction.

\subsection{Final Dataset Construction Approach}
Due to the limitations discussed in Section \ref{sec:initial-approach}, it was not feasible to label sufficiently large number of malware samples using a single unified approach. Specifically, the combination of strict API limits and the absence of the labels in the
VirusShare dataset made it impractical to label samples via VirusTotal and AVClass at the required scale.

As a result, the final dataset had to be constructed using binaries from multiple independent sources. This section describes the adopted approach, in which type-based and family-based datasets were created separately using different repositories. 

For type-based classification, unlabeled binaries were downloaded from VirusShare and later sorted based on their general behavioral characteristics. 

For family-based classification, binaries were obtained from VX Underground, which hosts monthly compressed archives originally exported from MalwareBazaar. These archives contain samples organized by submission date, and importantly, the filenames of the binaries explicitly include the corresponding malware family names
(e.g., \texttt{HEUR-Trojan-Banker.Win32.Qbot}). As a result, no additional labeling was required, and the dataset could be directly organized into malware families based on filename parsing.

\subsubsection{Type-Based Dataset}
\label{sec: type-based}

\begin{figure*}[t]
\centering
\begin{lstlisting}[caption=ClarAVy labels, style=malwarelistingstyle, label={lst:claravy}]
000251962a5d01c0d8bd79408744da13	53	  FILE:vbs|10
0000adda489b33d5f2d22d76e3bd8907    54    BEH:adware|6,BEH:downloader|6,PACK:nsis|1
0001bbaeafc6d232705939a859eda8dc     5    SINGLETON:0001bbaeafc6d232705939a859eda8dc
\end{lstlisting}
\end{figure*}

To construct the type-based dataset, malware binaries were obtained from the VirusShare repository. Since VirusShare does not provide any labeling, external labels were required to categorize the samples by malware type. Fortunately, precomputed behavior-based labels for VirusShare archives \texttt{00000} to \texttt{00465} were available through VX Underground, which redistributed data generated by the ClarAVy project. As shown in Listing \ref{lst:claravy}, each entry in the ClarAVy label file corresponds to one sample and contains its MD5 hash followed by a list of tags.

Each behavioral or structural tag is followed by a number indicating how many antivirus engines supported that specific label. According to Listing \ref{lst:claravy}, both the \texttt{adware} and \texttt{downloader} behaviors were supported by six engines each, while the sample itself received 54 positive detections in total according to VirusTotal scan result. Additionally, some tags such as \texttt{FILE:} or \texttt{PACK:} provide technical metadata unrelated to behavior, and some samples are marked as \texttt{SINGLETON}, indicating that no reliable classification could be assigned.

To prepare the dataset for further processing, the labels were filtered to exclude three specific groups: 
\begin{itemize} 
  \item Samples containing only non-behavioral tags (e.g., \texttt{FILE:} without any \texttt{BEH:} labels).
  \begin{itemize}
      \item A total of 4,238,221 samples were excluded.
    \end{itemize}
  
  \item Categories with low support - specifically, behavioral types with fewer than 1000 samples across all archives \texttt{00000} to \texttt{00465}.
    \begin{itemize}
      \item A total of 56 behavioral categories were excluded.
      \item These categories accounted for 11,206 samples in total.
    \end{itemize}

  \item Samples labeled as \texttt{SINGLETON}, which could not be reliably classified by ClarAVy.
  \begin{itemize}
      \item A total of 12,147,565 samples were excluded.
    \end{itemize}
\end{itemize}

After removing the excluded entries, the remaining hashes were grouped into malware type categories based on their associated \texttt{BEH:} tags. In total, 76 unique behavioral categories were identified, each representing a~distinct malware type used in further dataset construction.

It is important to note that all filtering and categorization steps described above were performed exclusively on precomputed behavior-based labels from ClarAVy in the form of plaintext files containing MD5 hashes followed by a~lists of tags, rather than on the binary files themselves. At this stage, no operations were conducted on the actual binary files. This design choice was motivated by performance considerations, as sorting and processing hash values is significantly faster and more scalable than handling large executable binaries.

For each behavioral category, a dedicated directory was created, containing individual label files for all relevant VirusShare archives in which samples of that type appeared. These files retained their original naming convention (e.g.,~\texttt{VirusShare\_00000\_labels.txt}) and stored only the hashes corresponding to the given category. This directory structure allowed for efficient retrieval and traceability of samples during the binary-level dataset assembly. 
%

\phantomsection
\label{par:label_distribution}
To facilitate further selection and practical construction of the dataset, a script was created to analyze the distribution of labeled samples across VirusShare archives for each behavioral category. This tool computes the number of MD5 hashes listed in every \texttt{VirusShare\_XXXXX\_labels.txt} file under each \texttt{BEH\_*} directory.\\

\noindent\textbf{Transition to Binary-Level Processing on External Server}\\
After the initial filtering and organization of metadata, the next phase of dataset construction involved working directly with the actual binary files. 

%


Actual binary files were extracted from the downloaded VirusShare archives and sorted into corresponding behavioral directories. In total, 14 behavioral categories were retained after filtering out categories with fewer than 1000 samples. Out of the original 76 behavioral categories identified, 17 contained no samples in the selected archives and were discarded immediately. 
The final selection is summarized in Table \ref{tab:behavioral_categories}.

\begin{table}[h]
\centering
\caption{Final selected behavioral categories with sample counts}
\begin{tabular}{|l|r|}
\hline
\textbf{Behavioral Type} & \textbf{Number of Samples} \\
\hline
BEH\_phishing   & 97,832 \\
BEH\_injector   & 13,600 \\
BEH\_virus      & 6,729 \\
BEH\_worm       & 6,416 \\
BEH\_downloader & 5,951 \\
BEH\_backdoor   & 5,010 \\
BEH\_coinminer  & 3,843 \\
BEH\_autorun    & 3,545 \\
BEH\_redirector & 2,834 \\
BEH\_dropper    & 1,728 \\
BEH\_iframe     & 1,530 \\
BEH\_spyware    & 1,262 \\
BEH\_clicker    & 1,247 \\
BEH\_adware     & 967 \\
\hline
\end{tabular}
\label{tab:behavioral_categories}
\end{table}

\subsubsection{Family-Based Dataset}
Unlike the type-based classification, constructing a dataset categorized by malware family proved to be significantly more challenging at the outset. Initially, no usable labels or structured metadata were available that would reliably map malware samples to their corresponding families. Attempts to extract this information from other sources such as VirusTotal or MalwareBazaar faced different issues. VirusTotal was limited by strict API constraints, while the MalwareBazaar API was not functioning reliably during the data collection phase, despite offering sufficient metadata in principle. 

The VX Underground platform had, in January 2025, begun redistributing large portions of the MalwareBazaar repository in the form of monthly archives. These archives are labeled by month (e.g., Bazaar.2025.03.7z) and contain samples submitted during the corresponding time period.

While this dataset structure is convenient for organizing and selecting samples, the publication schedule of these archives on VX Underground is not strictly periodic. As observed during this work, archives tend to appear with a delay of one to three months after the corresponding calendar month, but the exact timing of release is inconsistent.

The most recent archive included in this work corresponds to March 2025, making the family-based portion of the dataset relatively recent and well-suited for the analysis of contemporary malware families.\\

%

\noindent\textbf{Sorting Samples into Families}

Once the dataset had been collected, the next step was to organize the malware binaries into family-based categories. This was done by parsing the directory and filename structure of the samples extracted from the monthly archives. In most cases, the malware family name could be inferred by parsing the full filename of each sample, which often contains structured naming patterns (e.g., \texttt{UDS-Backdoor.Win32.Bladabindi.dej}). This enabled automated classification of binaries into families based on substring matching or regular expression extraction.

Before classification, an initial filtering step was applied to exclude samples that were not in the PE format. Only valid Windows PE files were retained, as the subsequent analysis pipeline was designed specifically for this file type. This filtering ensured format consistency and compatibility with tools used for feature extraction and classification. 


However, a number of binaries were labeled in a highly generic or inconsistent way, making it impossible to extract a reliable family name. These included samples with names such as \texttt{UDS-Trojan-Ransom.Win32.Generic}, which did not map to any meaningful or distinct malware family. Such entries were immediately excluded from further consideration to avoid ambiguity and noise in the resulting dataset.

Another issue was that not all samples could be reliably processed using automated parsing. Some filenames were too ambiguous or inconsistently structured to extract a meaningful family name without manual intervention. Since resolving these cases by hand was not practical at scale, such samples were excluded from the dataset.

Despite these limitations, the applied parsing process was able to extract family names for a significant portion of the dataset. Out of a total of 153,507 samples processed across all archives, 128,825 (84\%) were assigned a family name based on their filename structure. The remaining 24,682 samples did not meet the structural requirements for automated parsing and were excluded from further processing.

In total, this process yielded 65 distinct malware families, covering the most frequent and clearly identifiable classes found across the processed archives. Among the most prevalent were families such as 
\texttt{Agensla} (8,418~samples) and \texttt{Noon} (5,012 samples). 

As in the case of the Section \ref{sec: type-based} construction, the goal was to retain only those categories that were sufficiently populated to support meaningful analysis. Therefore, a minimum threshold of 1,000 samples per family was applied. Families that did not meet this criterion were excluded from the final dataset.

After applying the threshold of 1,000 samples per family, \textbf{17 malware families} remained in the dataset. These families represent the final output of the entire parsing and filtering process and were selected for use in the subsequent classification experiments. The final selection is summarized in Table~\ref{tab:final_family_counts}.

\begin{table}[h]
\centering
\caption{Final set of malware families with more than 1,000 samples}
\begin{tabular}{|l|r|}
\hline
\textbf{Family} & \textbf{Number of Samples} \\
\hline
Agensla       & 8,418  \\
Noon          & 5,012  \\
Taskun        & 4,888  \\
Convagent     & 3,155  \\
Stealerc      & 2,914  \\
Makoob        & 2,414  \\
Mokes         & 2,216  \\
Strab         & 2,191  \\
Remcos        & 2,043  \\
Androm        & 1,911  \\
Injuke        & 1,697  \\
SnakeLogger   & 1,595  \\
Crypt         & 1,434  \\
Zenpak        & 1,306  \\
Seraph        & 1,225  \\
Crysan        & 1,162  \\
DCRat         & 1,026  \\
\hline
\end{tabular}
\label{tab:final_family_counts}
\end{table}

The final dataset, which provides both feature-level and binary-level access to the samples from 14 malware types and 17 malware families, is publicly available on the \url{https://github.com/CS-and-AI/RawMal-TF}.

\section{Experiments}\label{Experiments}

This section presents a series of experiments aimed at evaluating the ability of machine-learning models to classify malware samples based on the datasets constructed in Section~\ref{Dataset Collection and Preparation}. The classification tasks target both the practical aspect of distinguishing malware from benign files and the structural complexity of differentiating between malware categories.

The experiments are divided into three parts. The first evaluates the ability to distinguish benign files from malware belonging to individual categories. The second focuses on binary discrimination between multiple malware types or families without including benign samples. Finally, the third explores a full multiclass classification setting, where samples are assigned to their respective malware categories without the presence of benign files.

All experiments rely on structured feature representations extracted directly from malware binaries. To ensure consistency and compatibility with the downstream processing tools, a uniform feature extraction pipeline was used throughout all classification tasks. The pipeline is based on the EMBER library, which defines a fixed feature schema and requires specific preprocessing steps. Adhering to this standardized process was necessary to produce valid input representations for model training.

Various machine-learning models were subsequently tested under controlled conditions, and their performance was evaluated using standard classification metrics such as accuracy, precision, recall, and F1-score.

All experiments were conducted on a dedicated computing platform equipped~with two Intel Xeon Gold 6136 CPUs (3.0 GHz, 12 cores each), an Nvidia Tesla P100 GPU (12 GB VRAM), and 754 GB of RAM, running Ubuntu 20.04.6 LTS.

\subsection{Feature extraction}
\label{sec:feature-extraction}

To enable machine learning-based classification, all binary samples were transformed into structured feature vector representations. For this purpose, the EMBER library was selected, as it provides a practical Python-based pipeline for processing Windows PE files and is widely used in the malware research community.

The original EMBER dataset (version 2018) contains a large number of benign and malicious samples in compressed form, where each sample is already represented by a pre-extracted feature vector \cite{2018arXiv180404637A}. 

Rather than using the precomputed data provided by EMBER, raw malware binaries collected from external sources (see Section~\ref{Dataset Collection and Preparation}) were processed independently using the EMBER feature extraction pipeline. Only benign samples used for training were taken directly from the original EMBER dataset, as they offer verified labels and structural consistency.

All extracted data were saved in the form of raw feature records. These were later converted into numerical vectors using custom processing scripts, as detailed in Subsection \ref{sec:custom-feature-pipeline}.

\subsubsection{Custom Feature Processing Pipeline}
\label{sec:custom-feature-pipeline}

This section describes the pipeline for transforming raw binaries into feature vectors.\\

\noindent\textbf{Step 1: Extracting Raw Features from Binary Samples}
\label{step1:extract-raw-features}

Each sample was processed using a custom Python script that automated batch extraction of static features from all classified binaries. The script iterated over directories containing malware samples grouped by type or family and applied the \texttt{PEFeatureExtractor()} class provided by the EMBER library to each executable file.

For every binary, the custom extraction script saved a dictionary containing three main fields:

\begin{itemize} 
\item \texttt{file} - the original filename of the binary sample, 
\item \texttt{raw\_features} - a structured dictionary of extracted static properties, 
\item \texttt{feature\_vector} - a numerical representation generated directly by the extractor. \end{itemize}

All three fields were initially saved during the extraction phase. However, only the \texttt{raw\_features} were used for further processing in this work. This decision was made in accordance with the internal design of the EMBER library, which expects \texttt{raw\_features} as input for its vectorization functions. As such, the precomputed \texttt{feature\_vector} field was ignored, and all numerical representations were subsequently derived by applying the EMBER vectorization logic directly to the \texttt{raw\_features}.

The \texttt{raw\_features} block consisted of the following 10 top-level components:

\begin{itemize}
\label{list:raw_feature_keys}
\item \texttt{sha256} - hash identifier of the binary, 
\item \texttt{histogram} - byte-value histogram of the entire file, 
\item \texttt{byteentropy} - entropy values over sliding windows across the file, 
\item \texttt{strings} - statistics derived from printable strings found in the file, 
\item \texttt{general} - basic file-level metadata such as size and compile timestamp, 
\item \texttt{header} - characteristics of the PE header, including machine type and subsystem, 
\item \texttt{section} - statistics aggregated over all PE sections (e.g., number, size, entropy), 
\item \texttt{imports} - list and count of imported libraries and functions, 
\item \texttt{exports} - list of exported symbols and their properties, 
\item \texttt{datadirectories} - presence and size of various PE data directories. 
\end{itemize}

These extracted components provided the foundation for all downstream processing. However, the raw feature dictionaries in their original form were not directly usable by the EMBER vectorization function. Additional adjustments were necessary to ensure structural compatibility and to supply required fields expected by the internal logic of the library. These modifications are described in the following step.\\

\noindent\textbf{Step 2: Preparing Raw Features for EMBER Vectorization}
\phantomsection
\label{step2:fix-raw-features}

Although the extracted \texttt{raw\_features} dictionaries contained most of the required static attributes, their overall structure did not fully match the expected input format used by the EMBER library. Specifically, the data were not directly compatible with the internal loading and preprocessing functions provided by the library, which assume the same JSON schema as used in the official EMBER 2018 dataset.

To ensure compatibility with the expected data format, four additional top-level metadata fields had to be added to each record. These fields are not produced during static feature extraction but are necessary for proper handling within the broader EMBER processing pipeline. Because this metadata was not available for the collected binaries, default placeholder values were used. The only exception was the \texttt{label} field, which was explicitly set to \texttt{1} for all samples to indicate that the dataset consisted exclusively of malware and did not include any benign files.

The following components were inserted into each record:

\begin{itemize}
\item \texttt{md5} - initialized as an empty string, since MD5 hashes were not computed,
\item \texttt{appeared} - set to an empty string as no submission date was available,
\item \texttt{label} - fixed to \texttt{1} for all samples to indicate the malware class,
\item \texttt{avclass} - initialized as an empty string, as AVClass labels were not available for either the type-based or family-based datasets.
\end{itemize}

After completing this step, each record contained all required top-level keys in a structure compatible with the expected EMBER schema. However, one final adjustment was necessary before proceeding to vectorization: the files had to be converted from standard \texttt{.json} format to \texttt{.jsonl} (JSON Lines) format, where each line represents one JSON-encoded sample. This is the format expected by EMBER's input handling functions, and without this transformation, the data could not be correctly parsed or processed by the subsequent steps in the pipeline.\\

\noindent\textbf{Step 3: Benign Sample Integration (Category vs. Benign)}
\label{step3:merge-clean}

As a result of the transformation described in \hyperref[step2:fix-raw-features]{Step~2}, each malware category (either type-based or family-based) was stored in a separate \texttt{.jsonl} file, where each line represented one binary sample described by its raw static features.

To prepare the datasets for supervised binary classification tasks, specifically those distinguishing malware from benign samples, a subset of these category-specific files was extended with benign executables. For each malware sample, one benign sample was randomly selected from the EMBER 2018 dataset, which provides feature vectors extracted from benign Windows binaries (label = 0). This resulted in balanced datasets with a 1:1 ratio of malicious to benign samples, suitable for evaluating detection performance.


All resulting datasets retained full compatibility with the EMBER processing pipeline, as the benign records used for augmentation were already in the expected \texttt{.jsonl} format with appropriate metadata and label structure.\\

\noindent\textbf{Step 4: Dataset Splitting and Final Formatting} 
\label{step4:splitting}

Once the datasets were finalized, whether they consisted solely of malware samples or included both malware and benign files, they were split into training and test sets and reformatted to comply with the input structure expected by the EMBER.

The EMBER library requires a specific directory layout and file-naming convention. Training data is expected to be divided across multiple files named in a fixed pattern (e.g., \texttt{train\_features\_0.jsonl}, \texttt{train\_features\_1.jsonl}, etc.), while all test data must be stored in a separate file named\\ \texttt{test\_features.jsonl}. 


To satisfy this requirement, each processed dataset was first randomly shuffled and then partitioned such that approximately 66\% of the records were used for training and the remaining 33\% for testing. The training portion was then evenly split into two files.

The resulting file sets were stored in separate directories corresponding to their respective malware category (e.g., \texttt{BEH\_worm/}, \texttt{Agensla/}), resulting in self-contained datasets that conform to the input structure used by EMBER's preprocessing and modeling utilities.\\


\noindent\textbf{Pipeline Summary}

This summary presents the complete pipeline for transforming raw binaries into machine-learning-ready feature vectors. All datasets were processed and stored in the expected format, allowing for seamless integration with the classification experiments.

\begin{enumerate}
\item \textbf{Step 1 - Feature Extraction:}  
  Static raw features were extracted from all classified binaries using the \texttt{PEFeatureExtractor()} class from the EMBER library.

\item \textbf{Step 2 - Schema Alignment:}  
  The extracted records were updated with required top-level metadata fields to ensure compatibility with EMBER's internal format.

\item \textbf{Step 3 - Benign Sample Integration:}  
  Benign samples from the EMBER 2018 dataset were added only to selected category-specific datasets to enable binary classification against benign files.

\item \textbf{Step 4 - Splitting and Formatting:}  
  All datasets were reshuffled and split into training and test files using EMBER-compliant filenames and directory structure.
\end{enumerate}

\subsection{Binary Classification Experiments}
Binary classification represents a fundamental task in malware detection, where the objective is to distinguish between malicious and benign samples. This section presents two categories of binary experiments: the first evaluates the ability to detect malware from benign files within a specific category (either a~malware type or family), while the second focuses on discriminating between different malware categories without including benign samples. These settings allow for an independent evaluation of how well individual categories can be detected and differentiated based solely on their static characteristics.\\

\noindent\textbf{Experimental Setup}

Each classification task was performed independently for every category (type or family) using the following standardized pipeline:

\begin{itemize}
   \item \textbf{Feature Standardization:} Feature vectors were first computed using the official EMBER vectorization pipeline, consisting of the functions \texttt{create\_vectorized\_features()} and \texttt{create\_metadata()}, which transform structured raw feature dictionaries into fixed-length numerical arrays. The resulting training and test vectors were then loaded via \texttt{read\_vectorized\_features()} and standardized using \texttt{StandardScaler} to ensure zero mean and unit variance across all features.

    \item \textbf{Feature Selection:} Univariate feature selection was applied using \texttt{SelectKBest} with the ANOVA F-score (\texttt{f\_classif}) as the scoring function \cite{scikitlearnSelectKBest}. The following values of $k$ were evaluated:
    \[
    k \in \{5, 10, 20, 30, 40, 50, 100\}
    \]
    The best value was selected based on validation accuracy using a baseline Random Forest model.
    
    \item \textbf{Model Evaluation:} The following classifiers were trained and optimized:
    \begin{itemize}
        \item \textbf{Random Forest}
        \begin{itemize}
            \item \texttt{n\_estimators}: \{100, 200\}
            \item \texttt{max\_depth}: \{None, 10, 20\}
        \end{itemize}
        
        \item \textbf{Support Vector Machine (SVM)}
        \begin{itemize}
            \item \texttt{kernel}: \{linear, rbf\}
            \item \texttt{C}: \{0.1, 1, 10\}
            \item \texttt{max\_iter}: 1000
        \end{itemize}
        
        \item \textbf{Multi-Layer Perceptron (MLP)}
        \begin{itemize}
            \item \texttt{hidden\_layer\_sizes}: \{(50,), (100,), (100, 50)\}
            \item \texttt{alpha}: \{0.0001, 0.001\}
            \item \texttt{max\_iter}: 1000
        \end{itemize}
        
        \item \textbf{K-Nearest Neighbors (KNN)}
        \begin{itemize}
            \item \texttt{n\_neighbors}: \{3, 5, 7\}
        \end{itemize}
        
        \item \textbf{XGBoost}
        \begin{itemize}
            \item \texttt{n\_estimators}: \{100, 200\}
            \item \texttt{max\_depth}: \{3, 6\}
            \item \texttt{eval\_metric}: \texttt{mlogloss}
        \end{itemize}
    \end{itemize}

    \item \textbf{Model Selection:} Each classifier was optimized using grid search with 3-fold cross-validation. The best model was selected based on validation accuracy and evaluated on the held-out test set.
    
    \item \textbf{Evaluation Metrics:} Performance was measured using:
    \begin{itemize}
        \item Accuracy - proportion of correct predictions,
        \item Precision - proportion of true positives among all predicted positives,
        \item Recall - proportion of true positives among all actual positives,
        \item F1-score - harmonic mean of precision and recall,
        \item Confusion matrix - matrix summarizing true vs. predicted classifications,
        \item Training time includes full model fitting with cross-validated hyperparameter search.
    \end{itemize}

\item \textbf{Dataset Composition:} In the \textit{category vs. benign} experiments, each dataset was constructed as a balanced binary classification task consisting of:
\begin{itemize}
    \item Malicious samples belonging to a specific category (either type- or family-based),
    \item An equal number of benign samples drawn from the EMBER 2018 dataset.
\end{itemize}
\end{itemize}
The procedure for creating these balanced datasets is described in Section~\ref{step3:merge-clean} of the feature processing pipeline. Each dataset was then split into training and test sets using a 66/33 ratio, with the training portion further divided into two parts to match the EMBER format.

In contrast, \textit{inter-category} experiments did not include any benign samples. These datasets consisted exclusively of malicious binaries belonging to two different categories, and were used to evaluate the discriminability between malware types or families.\\

%
%
%
%

\subsubsection{Malware Category vs. Benign}
The malware category vs. benign experiments were designed to evaluate how well machine learning models can distinguish malware from benign files when trained on samples from a specific malware category. This setting was applied independently to both the type-based and family-based datasets. Each classification task involved training and testing on a dataset composed of malware samples from a single category and an equal number of benign samples.

To assess how dataset size influences model performance, two configurations were tested:

\begin{itemize}
    \item \textbf{Full dataset:} Each experiment used all available malware samples for a given category, matched with an equal number of benign samples. This setting reflects the actual size and diversity of each category as summarized in Tables~\ref{tab:behavioral_categories} and~\ref{tab:final_family_counts}, providing models with the maximum amount of information per class.

    \item \textbf{Truncated dataset:} Each malware category was limited to 1,000 samples\footnote{With one exception - BEH\_adware category, which contained only 967 samples.}, paired with 1,000 benign samples. This setting allows for a uniform comparison across categories and evaluates model performance under limited data conditions. The truncated datasets are also used in subsequent experiments to ensure balanced category sizes and fair cross-category comparisons.

\end{itemize}

\noindent\textbf{Type-Based Detection (Full Scope)}\\
This section presents the results of binary classification experiments for each behavioral malware type using the full available dataset. 


\paragraph{Per-Model Performance Summary}
Across all behavioral types, Random Forest and XGBoost consistently delivered the best classification performance. Random Forest achieved the highest average precision (0.9942) and the lowest false positive rate (FPR = 0.0059), indicating strong specificity and minimal false alarms. It also achieved the highest average accuracy (0.9856) and exhibited the lowest variability (standard deviation of 0.0180).

XGBoost slightly outperformed Random Forest in terms of recall (0.9826) and false negative rate (FNR = 0.0174), making it particularly effective at identifying true malware samples. On the other end of the spectrum, KNN exhibited the highest FPR (0.0234) and lowest precision (0.9769), suggesting a tendency to misclassify benign samples.

\begin{figure*}[h]
    \centering
    \begin{minipage}[t]{0.48\textwidth}
        \centering
        \includegraphics[width=\textwidth]{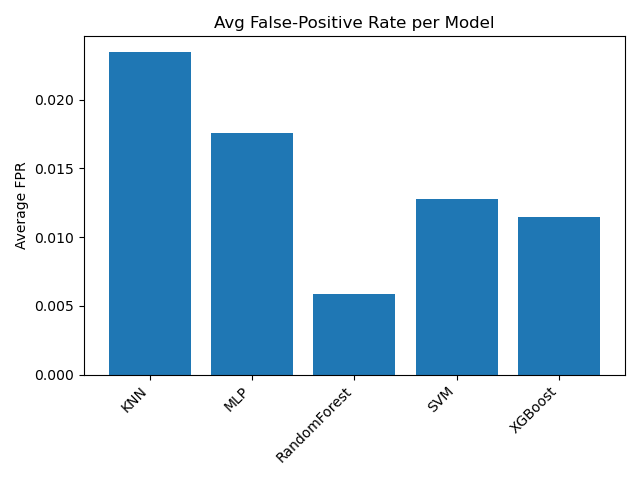}
        \caption{Average false positive rate (FPR) across all behavioral types.}
        \label{fig:average-fpr}
    \end{minipage}
    \hfill
    \begin{minipage}[t]{0.48\textwidth}
        \centering
        \includegraphics[width=\textwidth]{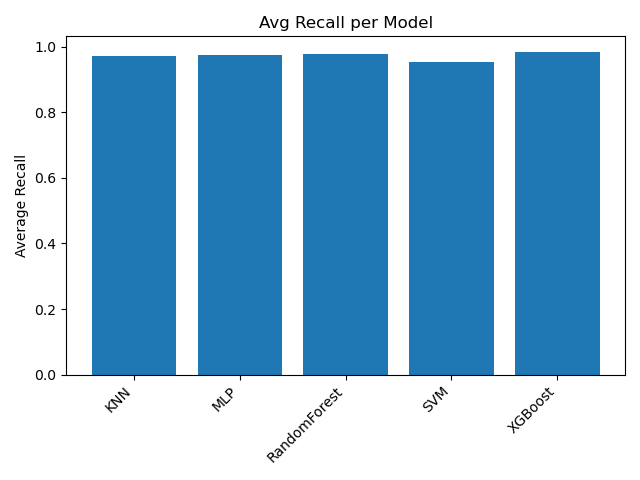}
        \caption{Average recall across all behavioral types.}
        \label{fig:average-recall}
    \end{minipage}

    \vspace{1em} 

    \begin{minipage}[t]{0.48\textwidth}
        \centering
        \includegraphics[width=\textwidth]{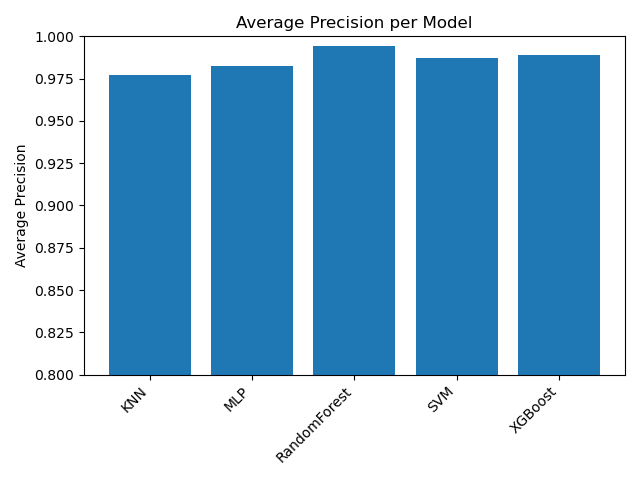}
        \caption{Average precision across all behavioral types.}
        \label{fig:average-precision}
    \end{minipage}
    \hfill
    \begin{minipage}[t]{0.48\textwidth}
        \centering
        \includegraphics[width=\textwidth]{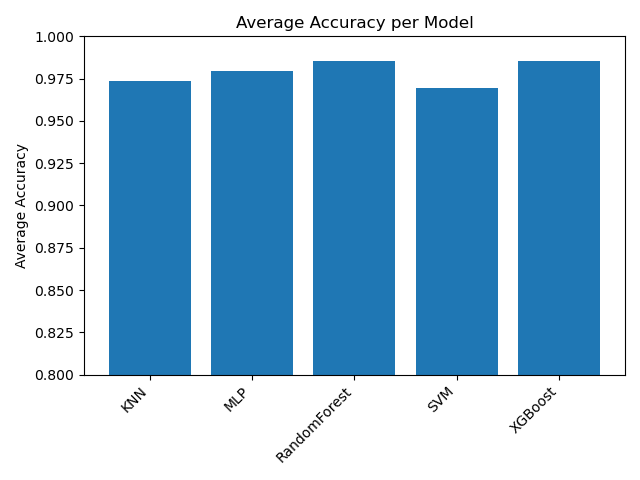}
        \caption{Average accuracy across all behavioral types.}
        \label{fig:average-accuracy}
    \end{minipage}
\end{figure*}

Figures~\ref{fig:average-fpr}, \ref{fig:average-recall}, \ref{fig:average-precision}, and \ref{fig:average-accuracy} summarize these trends graphically, showing the average FPR, recall, precision, and accuracy per model.

\paragraph{Trade-off Between Speed and Performance}
Table~\ref{tab:tradeoff-fpr-speed} presents a trade-off analysis comparing each model's average accuracy, training time, and stability, where stability is measured by the standard deviation. While KNN is by far the fastest ($0.29\text{s}$), it also ranks lowest in both accuracy and stability. MLP provides decent accuracy but suffers from long training times ($5.6\text{s}$). In contrast, XGBoost offers an excellent balance of high accuracy, low FPR, and fast execution ($0.38\text{s}$), while Random Forest achieves the top scores in most performance metrics, albeit at higher training cost.

\begin{table*}[h]
\centering
\resizebox{\textwidth}{!}{%
\begin{tabular}{|l|c|c|c|c|}
\hline
\textbf{Model} & \textbf{Avg Accuracy} & \textbf{Std Accuracy} & \textbf{Avg Training Time (s)} & \textbf{Performance Highlights} \\
\hline
KNN           & 0.9735 & 0.0295 & 0.29 & Speed only \\
MLP           & 0.9792 & 0.0230 & 5.55 & Moderate accuracy \\
SVM           & 0.9692 & 0.0372 & 0.50 & Low precision, unstable \\
XGBoost       & 0.9855 & 0.0180 & 0.38 & Balanced performance \\
RandomForest  & 0.9856 & 0.0180 & 2.09 & Best overall performance \\
\hline
\end{tabular}%
}
\caption{Model trade-off: accuracy, standard deviation of the accuracy, and training time (behavioral types).}
\label{tab:tradeoff-fpr-speed}
\end{table*}

\paragraph{Category-Level Observations}
Some malware types proved easier to classify than others. \texttt{BEH\_iframe}, \texttt{BEH\_phishing}, and \texttt{BEH\_redirector} achieved perfect classification (100\% accuracy) across all models. Conversely,\\ \texttt{BEH\_spyware} emerged as the most challenging, with an average accuracy of 92.6\% and the highest FPR (10.1\%), particularly for KNN.

This discrepancy suggests that certain behavioral types (e.g., redirectors or phishing droppers) have more distinctive static characteristics, while others (like spyware) may share features with benign executables, complicating detection.

\paragraph{Summary}
The full-dataset experiments on behavioral types reveal that most categories are highly separable from benign samples using static features, with multiple models achieving near-perfect performance. However, model choice significantly affects performance in edge cases, particularly for more ambiguous behaviors like spyware or adware. XGBoost and Random Forest emerged as the most robust classifiers, striking a strong balance between accuracy, training time, and reliability across types.\\

\noindent\textbf{Family-Based Detection (Full Scope)} \\
This section presents the results of binary classification experiments conducted on malware families using the full dataset available for each family. 

\paragraph{Per-Model Performance Summary} Across all malware families, Random Forest and XGBoost again emerged as the top-performing classifiers. Random Forest achieved the highest average precision (0.9936) and the second-lowest false positive rate (FPR = 0.0065), indicating high specificity and reliability. It also recorded one of the highest average accuracies (0.9891) with low variability (standard deviation = 0.0109).

XGBoost delivered the best overall accuracy (0.9898) and recall (0.9878), along with the lowest false negative rate (FNR = 0.0122), making it especially effective in detecting malware samples with minimal missed detections. In contrast, KNN exhibited the weakest precision (0.9661) and the highest FPR (0.0346), confirming its tendency to overpredict malware labels, which can lead to an increased number of false alarms.

\begin{figure*}[h]
    \centering
    \begin{minipage}[t]{0.48\textwidth}
        \centering
        \includegraphics[width=\textwidth]{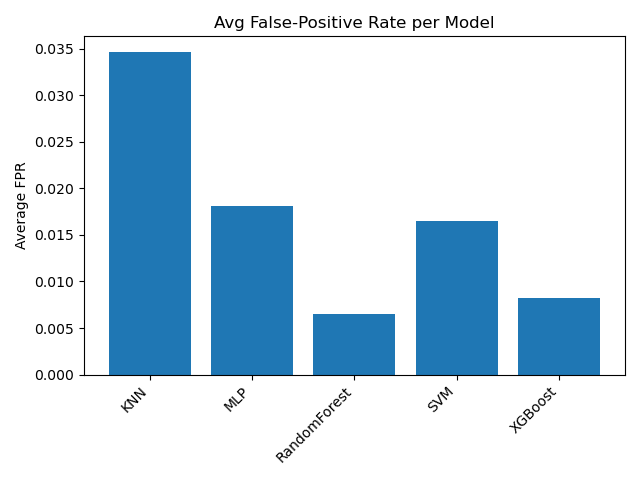}
        \caption{Average false positive rate (FPR) across all malware families.}
        \label{fig:average-fpr-fam}
    \end{minipage}
    \hfill
    \begin{minipage}[t]{0.48\textwidth}
        \centering
        \includegraphics[width=\textwidth]{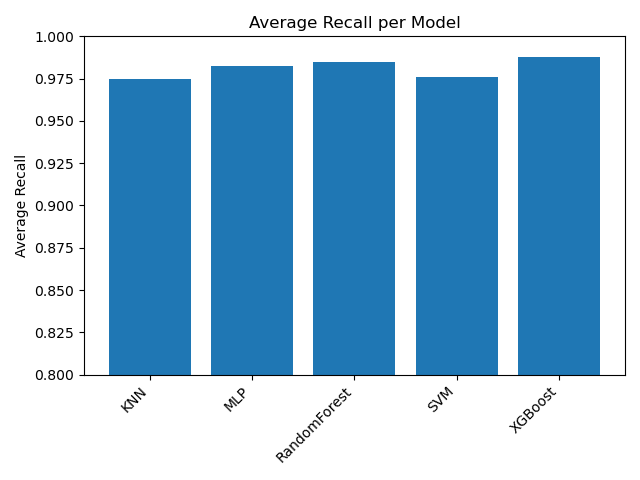}
        \caption{Average recall across all malware families.}
        \label{fig:average-recall-fam}
    \end{minipage}

    \vspace{1em} 

    \begin{minipage}[t]{0.48\textwidth}
        \centering
        \includegraphics[width=\textwidth]{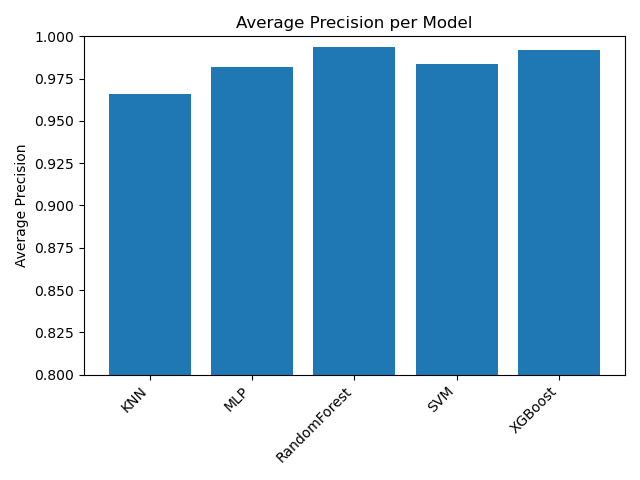}
        \caption{Average precision across all malware families.}
        \label{fig:average-precision-fam}
    \end{minipage}
    \hfill
    \begin{minipage}[t]{0.48\textwidth}
        \centering
        \includegraphics[width=\textwidth]{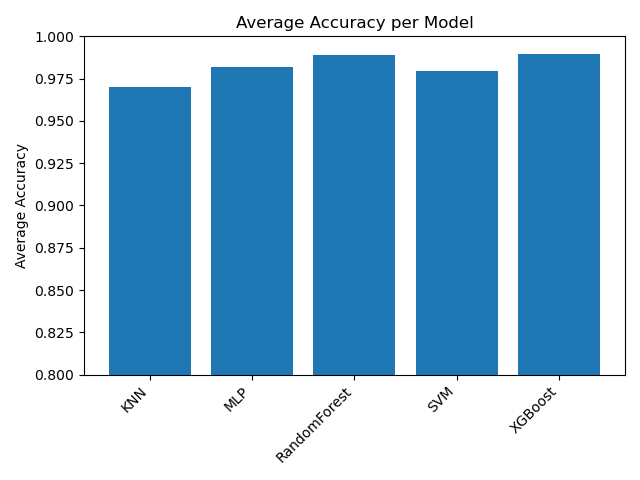}
        \caption{Average accuracy across all malware families.}
        \label{fig:average-accuracy-fam}
    \end{minipage}
\end{figure*}

The average performance metrics across all malware families are visualized in Figures~\ref{fig:average-fpr-fam}, \ref{fig:average-recall-fam}, \ref{fig:average-precision-fam}, and \ref{fig:average-accuracy-fam}, providing an overview of FPR, recall, precision, and accuracy per classifier.

\paragraph{Trade-off Between Speed and Performance} Table~\ref{tab:tradeoff-fpr-speed-fam} summarizes the trade-off between classification performance, training time, and model stability across all family-based experiments. KNN remained the fastest model by a~wide margin, with an average training time of just 0.12s. However, this came at the cost of reduced performance (average accuracy 97.02\%) and higher variance in results.

In contrast, XGBoost and Random Forest achieved nearly identical top-level accuracy (98.98\% and 98.91\% respectively), with XGBoost offering substantially faster training (0.41s vs. 4.19s) and slightly better consistency (lowest standard deviation in both accuracy and F1-score). MLP performed well overall, but suffered from high training times (4.46s). SVM showed balanced performance but remained less stable and more error-prone than the leading ensemble models.

\begin{table*}[h]
\centering
\caption{Model trade-off: accuracy, standard deviation of the accuracy, and training time (malware families).}
\label{tab:tradeoff-fpr-speed-fam}
\resizebox{\textwidth}{!}{%
\begin{tabular}{|l|c|c|c|c|}
\hline
\textbf{Model} & \textbf{Avg Accuracy} & \textbf{Std Accuracy} & \textbf{Avg Training Time (s)} & \textbf{Performance Highlights} \\
\hline
KNN           & 0.9702 & 0.0189 & 0.12 & Speed only \\
MLP           & 0.9820 & 0.0112 & 4.46 & Moderate accuracy \\
SVM           & 0.9796 & 0.0128 & 0.59 & Balanced but unstable \\
XGBoost       & 0.9898 & 0.0091 & 0.41 & Best overall performance \\
RandomForest  & 0.9891 & 0.0109 & 4.19 & High accuracy, less efficient \\
\hline
\end{tabular}%
}
\end{table*}

\paragraph{Category-Level Observations}
Classification difficulty varied significantly across malware families. \texttt{Convagent} proved to be the most challenging, with the lowest average accuracy (95.97\%) among all categories. This result may be attributed to weaker discriminative characteristics within the family or greater structural similarity to benign samples. Additionally, \texttt{Androm} exhibited the highest false positive rate (FPR = 0.0883) when classified by KNN, indicating frequent misclassification of benign files.

In contrast, families like \texttt{Makoob}, \texttt{Mokes}, and \texttt{Agensla} achieved near-perfect classification accuracy (above 99.5\%), suggesting that these families contain distinctive patterns easily separable from benign files. In some cases, such as \texttt{SnakeLogger}, \texttt{Crysan}, and \texttt{Injuke}, multiple classifiers even achieved a zero false positive rate, demonstrating excellent specificity.

\paragraph{Summary} The full-dataset experiments on malware families demonstrate that family-based detection yields consistently strong classification results across all evaluated models. Compared to type-based classification, family-level tasks generally achieved higher average accuracy and lower variability, reflecting the greater internal consistency and semantic cohesion of malware families.

XGBoost and Random Forest emerged as the top-performing classifiers, achieving near-identical average accuracy (98.9\%), with Random Forest exhibiting an average false positive rate of about 0.65\% and recall of roughly 98.48\%, and XGBoost an average false positive rate of about 0.82\% and recall of roughly 98.78\%. KNN, although computationally efficient, again lagged behind in absolute performance and exhibited the highest false positive rate (3.46\%). \\

\noindent\textbf{Type-Based Detection (Truncated Scope)}\\
This section evaluates binary classification performance for malware types using the truncated dataset configuration. Each type-specific dataset contains approximately 1,000 malicious and 1,000 benign samples, enabling a uniform and controlled comparison across categories. The same experimental pipeline was applied as in previous sections, and the results are analyzed using the standard evaluation metrics. 

\paragraph{Per-Model Performance Summary}
Across all malware types in the truncated dataset setting, Random Forest and XGBoost again demonstrated top-tier classification performance. Random Forest achieved the highest average precision (0.9899), the highest average F1-score (0.9757), and the lowest false positive rate \mbox{(FPR = 0.0103)}. It also maintained the best average accuracy (97.57\%).

XGBoost closely followed with an average accuracy of $97.51\%$ and the highest recall ($96.98\%$), indicating strong sensitivity in identifying malicious samples. It also showed the lowest variability in accuracy (standard deviation $0.0268$), making it the most stable model across categories.

In contrast, KNN, while very fast to train, exhibited the lowest performance metrics in nearly every aspect, including the lowest average recall (0.9600) and the highest false positive rate \mbox{(FPR = 0.0296)}, reinforcing earlier observations from the full-scope experiment.

\begin{figure*}[h]
    \centering
    \begin{minipage}[t]{0.48\textwidth}
        \centering
        \includegraphics[width=\textwidth]{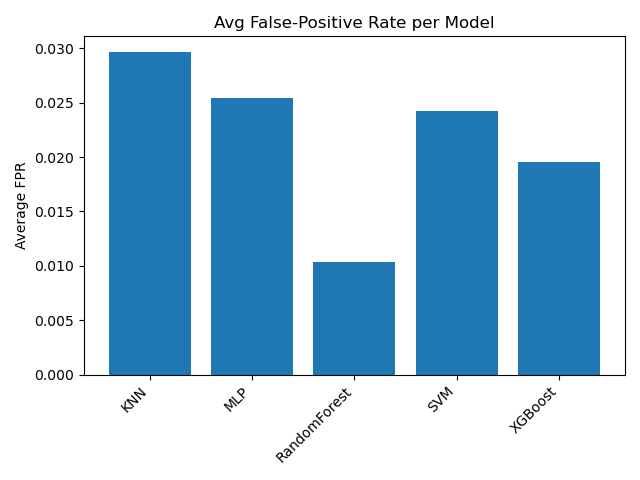}
        \caption{Average false positive rate (FPR) across truncated malware types.}
        \label{fig:average-fpr-types-trunc}
    \end{minipage}
    \hfill
    \begin{minipage}[t]{0.48\textwidth}
        \centering
        \includegraphics[width=\textwidth]{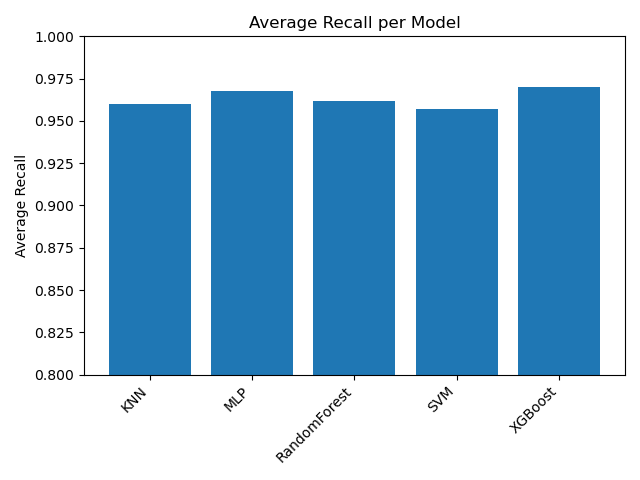}
        \caption{Average recall across truncated malware types.}
        \label{fig:average-recall-types-trunc}
    \end{minipage}

    \vspace{1em} 
    
    \begin{minipage}[t]{0.48\textwidth}
        \centering
        \includegraphics[width=\textwidth]{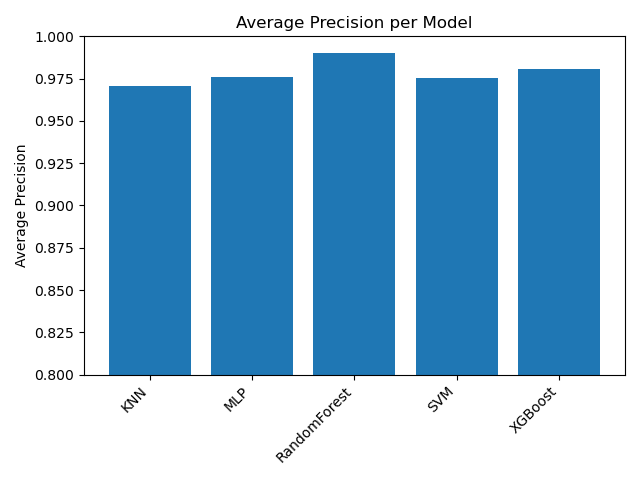}
        \caption{Average precision across truncated malware types.}
        \label{fig:average-precision-types-trunc}
    \end{minipage}
    \hfill
    \begin{minipage}[t]{0.48\textwidth}
        \centering
        \includegraphics[width=\textwidth]{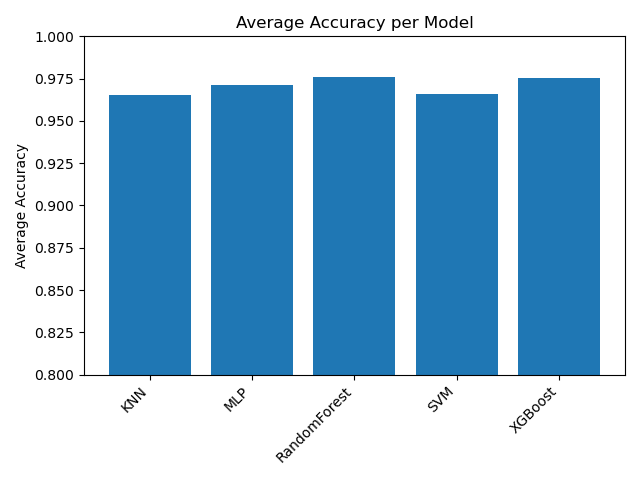}
        \caption{Average accuracy across truncated malware types.}
        \label{fig:average-accuracy-types-trunc}
    \end{minipage}
\end{figure*}

Figures~\ref{fig:average-fpr-types-trunc}, \ref{fig:average-recall-types-trunc}, \ref{fig:average-precision-types-trunc}, and \ref{fig:average-accuracy-types-trunc} provide a visual summary of model performance across the truncated malware type datasets.

\paragraph{Trade-off Between Speed and Performance}
Table~\ref{tab:tradeoff-fpr-speed-trunc} presents the trade-off analysis for the truncated type-based datasets. As in the full-scope setting, KNN remains the fastest model (average training time $0.03\,\text{s}$), but also exhibits the lowest accuracy ($0.9651$) and the highest variability in results (standard deviation $0.0342$). This reinforces its position as a speed-focused baseline rather than a high-precision classifier.

XGBoost and Random Forest emerged as the two strongest classifiers in the truncated type-based setting, delivering nearly identical average accuracy. XGBoost demonstrated the strongest stability, exhibiting the lowest variance in accuracy ($0.0268$) along with the fastest training time ($0.20\,\mathrm{s}$). Random Forest, by contrast, delivered the highest precision ($98.99\%$) and the lowest false positive rate ($1.03\%$), but required over three times longer to train ($0.71\,\mathrm{s}$) and showed slightly greater variability.

\begin{table*}[h]
\centering
\resizebox{\textwidth}{!}{%
\begin{tabular}{|l|c|c|c|c|}
\hline
\textbf{Model} & \textbf{Avg Accuracy} & \textbf{Std Accuracy} & \textbf{Avg Training Time (s)} & \textbf{Performance Highlights} \\
\hline
KNN           & 0.9651 & 0.0342 & 0.03 & Speed only \\
MLP           & 0.9713 & 0.0289 & 1.65 & Moderate accuracy \\
SVM           & 0.9661 & 0.0339 & 0.09 & The most unstable \\
XGBoost       & 0.9751 & 0.0268 & 0.20 & Best overall trade-off \\
RandomForest  & 0.9757 & 0.0283 & 0.71 & Slightly higher accuracy, but slower \\
\hline
\end{tabular}%
}
\caption{Model trade-off: accuracy, standard deviation of the accuracy, and training time (truncated behavioral types).}
\label{tab:tradeoff-fpr-speed-trunc}
\end{table*}

\paragraph{Category-Level Observations}
As in the full-scope experiments, individual malware types exhibited varying levels of classification difficulty. The easiest categories remained \texttt{BEH\_phishing} and \texttt{BEH\_redirector}, which achieved perfect classification (100\% accuracy) even under the truncated dataset constraint. These types likely retain distinctive static features that remain identifiable despite reduced training data. \texttt{BEH\_iframe} followed closely, with near-perfect accuracy (99.94\%), confirming its ease of detection across models.

Conversely, \texttt{BEH\_spyware} consistently proved the most challenging, reaching only an average accuracy of 90.9\%. This category also recorded the highest false positive rate in the truncated experiments, particularly under the MLP classifier (FPR = 11.4\%). These results are in line with the full-scope findings and confirm that spyware samples share more structural similarities with benign files, making them more prone to misclassification under limited data.

Compared to the full dataset, all categories experienced a slight drop in accuracy due to the reduced number of samples. However, despite the reduced dataset size, the relative ranking of category difficulty remained largely consistent with the full-scope experiments. This stability suggests that even with fewer training samples, the truncated datasets preserved enough representative features to support reliable classification.

%
%
%
%

\paragraph{Summary}
The truncated dataset experiments confirm that reliable malware detection remains feasible under constrained data conditions. Although all models experienced a slight decrease in performance compared to the full-scope setting, the relative ranking of classifiers remained stable. XGBoost continued to offer the best trade-off between accuracy and efficiency, while Random Forest maintained the highest precision at a modestly increased computational cost.

Category-level patterns were largely preserved, with \texttt{BEH\_phishing} and \texttt{BEH\_redirector} still achieving perfect classification, and \texttt{BEH\_spyware} remaining the most difficult to detect. 

Overall, the truncated dataset provided a valid framework for controlled comparisons across malware types, though full-scope datasets better capture the complexity of real-world detection scenarios.\\

\noindent\textbf{Family-Based Detection (Truncated Scope)}\\
This section presents binary classification results for malware families using balanced truncated datasets of 1,000 malware and 1,000 benign samples. The experiments followed the same pipeline as before, and results are compared with full-scope experiments to assess the impact of dataset size reduction. 

\paragraph{Per-Model Performance Summary} Across all malware families under truncated data conditions, XGBoost and Random Forest remained the top-performing classifiers. XGBoost achieved the highest average accuracy (0.9866), recall (0.9872), along with the lowest false negative rate (FNR = 0.0128) and the lowest false positive rate (FPR = 0.0141). It also exhibited the lowest variability (standard deviation $0.0171$), confirming its consistency and robustness across all families.

Random Forest followed closely, with slightly lower accuracy (0.9855), and precision (0.9840). Its FPR remained low at 0.0161, with a marginally higher false negative rate (FNR = 0.0130). Despite these slight differences, both models proved highly effective under limited data.

On the other hand, KNN showed the weakest precision (0.9694) and the highest FPR (FPR = 0.0307), confirming its relative inefficiency and tendency to produce false positives. MLP and SVM performed moderately well, with MLP offering balanced metrics but higher training cost, and SVM delivering stable but slightly lower overall accuracy.

\begin{figure*}[h]
    \centering
    \begin{minipage}[t]{0.48\textwidth}
        \centering
        \includegraphics[width=\textwidth]{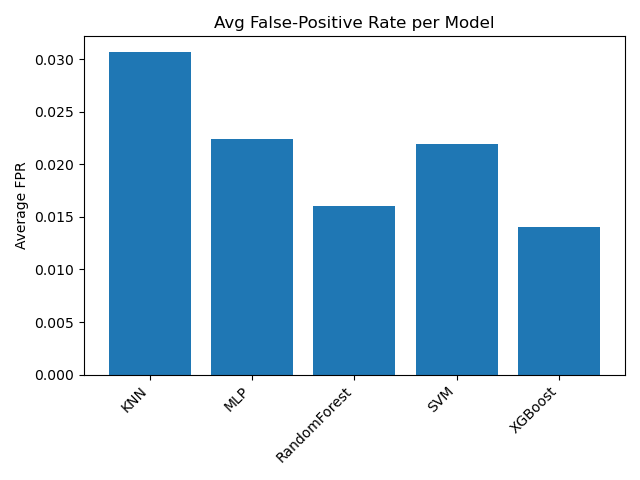}
        \caption{Average false positive rate (FPR) across truncated malware families.}
        \label{fig:average-fpr-fam-trun}
    \end{minipage}
    \hfill
    \begin{minipage}[t]{0.48\textwidth}
        \centering
        \includegraphics[width=\textwidth]{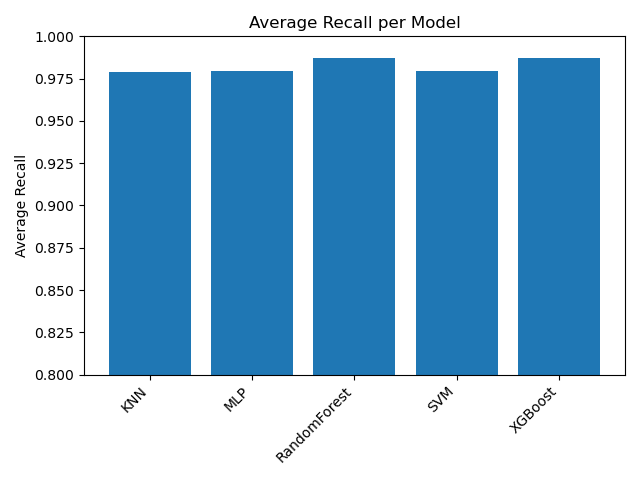}
        \caption{Average recall across truncated malware families.}
        \label{fig:average-recall-fam-trun}
    \end{minipage}

	\vspace{1em} 
	
    \begin{minipage}[t]{0.48\textwidth}
        \centering
        \includegraphics[width=\textwidth]{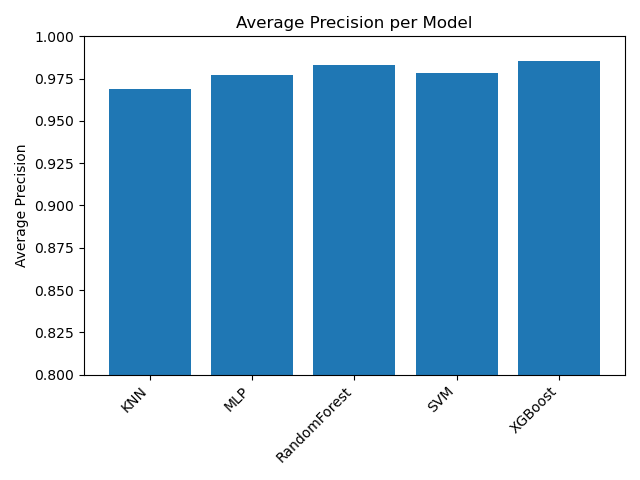}
        \caption{Average precision across truncated malware families.}
        \label{fig:average-precision-fam-trun}
    \end{minipage}
    \hfill
    \begin{minipage}[t]{0.48\textwidth}
        \centering
        \includegraphics[width=\textwidth]{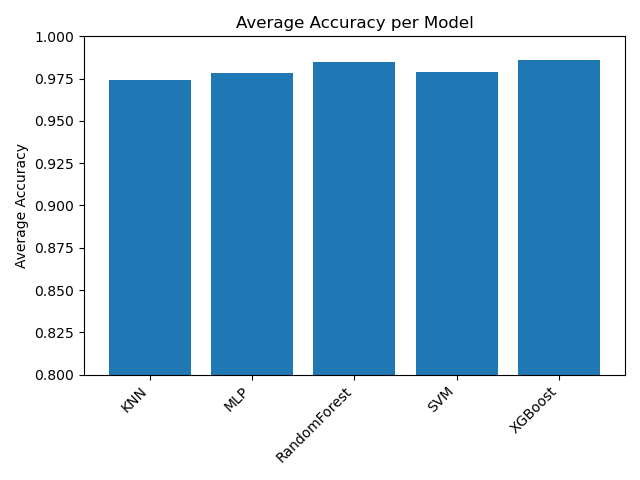}
        \caption{Average accuracy across truncated malware families.}
        \label{fig:average-accuracy-fam-trun}
    \end{minipage}
\end{figure*}

The average performance metrics across all malware families are visualized in Figures~\ref{fig:average-fpr-fam-trun}, \ref{fig:average-recall-fam-trun}, \ref{fig:average-precision-fam-trun}, and \ref{fig:average-accuracy-fam-trun}, providing an overview of FPR, recall, precision, and accuracy per classifier.

\paragraph{Trade-off Between Speed and Performance} Table~\ref{tab:tradeoff-fpr-speed-fam-trun} presents the trade-off analysis for truncated family-based classification. XGBoost delivered the best balance between accuracy (0.9866), low variability (Std $=$ 0.0171), and fast training time (0.20 s), making it the most effective choice overall. Random Forest offered nearly identical accuracy (0.9855) but required significantly more time to train (0.82 s) and showed slightly higher variability. KNN remained the fastest option, but with the lowest accuracy (0.9735) and stability. MLP and SVM achieved moderate results, with SVM being more efficient computationally.

\begin{table*}[h]
\centering
\resizebox{\textwidth}{!}{%
\begin{tabular}{|l|c|c|c|l|}
\hline
\textbf{Model}    & \textbf{Avg Accuracy} & \textbf{Std Accuracy} & \textbf{Avg Training Time (s)} & \textbf{Performance Highlights}                           \\
\hline
KNN               & 0.9735               & 0.0187               & 0.03                          & Speed only                                  \\
MLP               & 0.9784               & 0.0190               & 1.70                          & Balanced metrics, high cost                 \\
SVM               & 0.9791               & 0.0178               & 0.08                          & Efficient with moderate accuracy            \\
XGBoost           & 0.9866               & 0.0171               & 0.20                          & Best overall performance                    \\
Random Forest     & 0.9855               & 0.0189               & 0.82                          & High precision, slightly slower             \\
\hline
\end{tabular}%
}
\caption{Model trade-off: accuracy, standard deviation of the accuracy, and training time (malware families, truncated).}
\label{tab:tradeoff-fpr-speed-fam-trun}
\end{table*}

\paragraph{Category-Level Observations}  
The truncated dataset experiments confirmed that certain malware families remain easier to classify, even with limited data. \texttt{Makoob}, \texttt{Agensla}, and \texttt{Taskun} achieved the highest average accuracies ($\geq 99.6\%$), demonstrating their strong and distinctive static signatures. Conversely, \texttt{Convagent} posed the greatest challenge, with an average accuracy of $93.40\%$ and the highest false positive rate (FPR = $9.76\%$) for KNN.

Several families showed improved classification performance under the truncated configuration. For example, \texttt{Crysan} improved from $96.8\%$ to $98.71\%$, and \texttt{SnakeLogger} slightly increased from $98.2\%$ to $98.80\%$. 

Overall, while a few families experienced minor accuracy drops, others benefited from the more constrained data, and the results confirm that static feature-based family classification remains reliable even under data limitations.

\paragraph{Summary} The truncated-scope experiments on malware families confirmed that classification performance remained high even under reduced data conditions. XGBoost consistently emerged as the top-performing model, achieving the highest average accuracy ($0.9866$), and recall, along with the lowest false positive and false negative rates. Random Forest followed closely, with nearly identical performance but higher training time.

When comparing to the full-scope results, the overall classification accuracy remained stable, with only a marginal drop observed across models. Interestingly, some families like \texttt{Makoob} and \texttt{Agensla} continued to achieve near-perfect accuracy, demonstrating the robustness of their distinctive static features. However, families such as \texttt{Convagent} and \texttt{Androm} remained more challenging, with increased false positive rates in specific model configurations.


\begin{table*}[h]
\centering

\begin{tabular}{|l|c|c|c|}
\hline
\textbf{Model} & \textbf{Avg Accuracy} & \textbf{Std Accuracy} & \textbf{Avg Training Time (s)} \\
\hline
KNN           & 0.9663 & 0.0707  & 0.0395 \\
MLP           & 0.9722 & 0.0503   & 3.7977 \\
SVM           & 0.9700 & 0.0490  & 0.0719 \\
XGBoost       & 0.9739 & 0.0529  & 20.5035 \\
RandomForest  & 0.9750 & 0.0510  & 0.7916 \\
\hline
\end{tabular}%

\caption{Summary of average model performance in interclass type-based experiments.}
\label{tab:interclass-summary}
\end{table*}

\subsubsection{Interclass Malware Classification}
In contrast to the category vs. benign experiments, where the primary task was to distinguish malware from benign software, this section focuses on interclass malware classification, i.e., differentiating between two distinct malware categories. The objective is to evaluate how effectively machine learning models can discriminate between malware types or families based solely on their static characteristics.

Each experiment in this section involves binary classification where the task is to distinguish between two malware categories, such as type-based comparisons (e.g., \texttt{BEH\_worm} vs. \texttt{BEH\_backdoor}) or family-based comparisons (e.g., \texttt{Remcos} vs. \texttt{Androm}). 

For clarity in reporting metrics (error rate, precision, and recall), we follow the naming convention \texttt{<ClassA>\_0\_<ClassB>\_1}, where the suffix "\_1" (i.e., \texttt{ClassB}) denotes the positive class. Consequently, all binary metrics are computed with respect to \texttt{ClassB} as the positive class. \\


\noindent\textbf{Classification Inter-Types}
\paragraph{Per-Model Performance Summary}
Across all inter-type classification tasks, Random Forest and XGBoost demonstrated the most robust performance, achieving the highest average accuracy ($97.5\%$ and $ 97.4\%$, respectively). MLP followed closely with a strong average accuracy of $97.2\%$, while SVM and KNN lagged slightly behind at $97.0\%$ and $96.6\%$ accuracy, respectively.

Notably, KNN, despite being the fastest model (average training time $0.04~\text{s}$), exhibited the highest variability in performance (standard deviation of accuracy $0.07$) and the highest error rate ($0.033$), indicating that its ability to differentiate between similar malware types is limited compared to more sophisticated models. MLP showed competitive precision and recall but required significantly more training time ($3.8 \text{s}$), while Random Forest offered the best balance between accuracy, consistency, and efficiency.

In Table~\ref{tab:interclass-summary}, the average accuracy, error rate, and training time are computed over all pairs of malware categories.

\begin{table}[h]
\centering
\resizebox{\columnwidth}{!}{%
\begin{tabular}{|l|c|}
\hline
\textbf{Category Pair} & \textbf{Misclassification Rate} \\
\hline
autorun $\leftrightarrow$ worm & 20\% (autorun), 44.5\% (worm) \\
injector $\leftrightarrow$ downloader & 41.2\% (injector), 20.8\% (downloader) \\
virus $\leftrightarrow$ worm & 32.2\% (virus), 8.1\% (worm) \\
\hline
redirector $\leftrightarrow$ worm         & 0.0\% \\  
phishing $\leftrightarrow$ worm           & 0.0\% \\  
injector $\leftrightarrow$ redirector     & 0.0\% \\  
injector $\leftrightarrow$ phishing       & 0.0\% \\  
iframe $\leftrightarrow$ worm             & 0.0\% \\  
iframe $\leftrightarrow$ injector         & 0.0\% \\  
autorun $\leftrightarrow$ redirector      & 0.0\% \\  
autorun $\leftrightarrow$ phishing        & 0.0\% \\  
autorun $\leftrightarrow$ clicker         & 0.0\% \\  
autorun $\leftrightarrow$ iframe          & 0.0\% \\  
\hline
\end{tabular}
}
\caption{Misclassification rates for selected type pairs in interclass classification.}
\label{tab:type-separability}
\end{table}

\paragraph{Type Separability Observations}
The analysis revealed that certain malware types are inherently more challenging to distinguish. For instance, autorun and worm types exhibited high mutual confusion, with 20\% of autorun samples misclassified as worm and 44.5\% of worm samples as autorun. Similarly, injector and downloader also showed significant overlap, as detailed in Table~\ref{tab:type-separability}.

In contrast, several type pairs were perfectly separable by all models. A total of 10 pairs, such as redirector vs. worm, phishing vs. worm, and injector vs. redirector, achieved zero misclassifications across all models, highlighting strong static differences in their features.

These findings underscore that while some behavioral types share similar traits, others possess distinct enough characteristics to enable highly accurate classification, even without benign samples in the dataset.

\begin{figure}

        \includegraphics[width=\columnwidth]{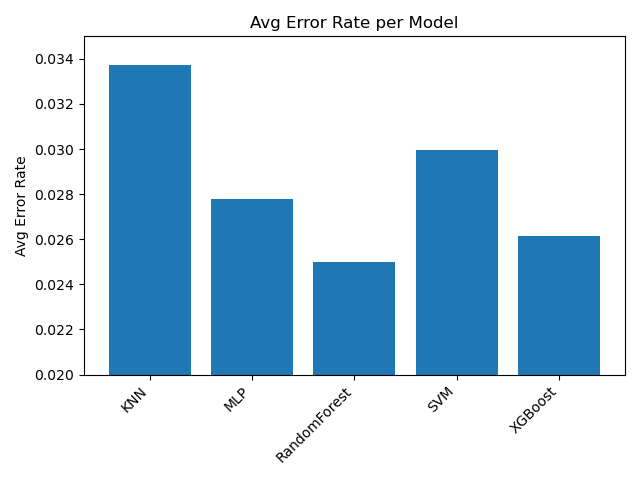}
        \caption{Average Error Rate across all inter-type tasks.}
        \label{fig:interclass-fpr}

        \includegraphics[width=\columnwidth]{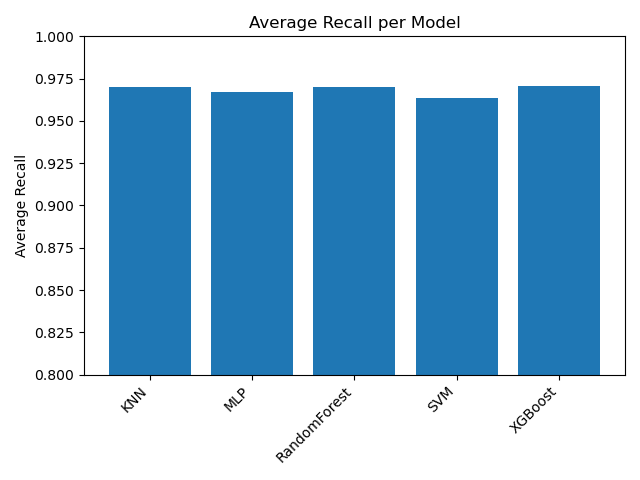}
        \caption{Average recall across all inter-type tasks.}
        \label{fig:interclass-recall}

        \includegraphics[width=\columnwidth]{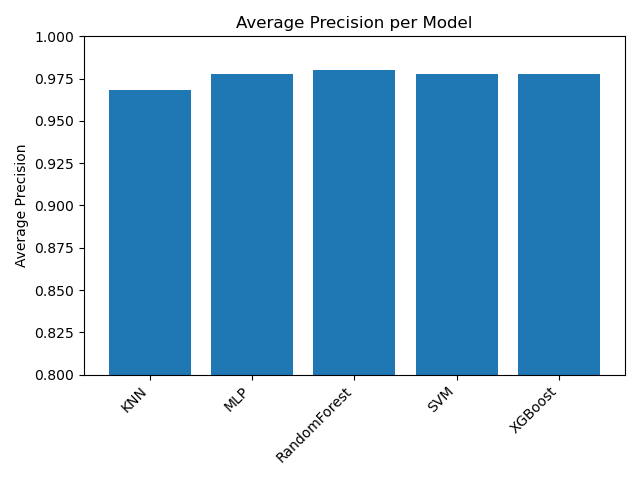}
        \caption{Average precision across all inter-type tasks.}
        \label{fig:interclass-precision}
\end{figure}

Figure~\ref{fig:interclass-fpr} summarizes the average error rate per model, while Figures~\ref{fig:interclass-recall} and~\ref{fig:interclass-precision} illustrate differences in recall and precision. These visualizations follow the same structure as in the previous classification experiments, providing a consistent comparative view of model behavior across different experimental settings.

\paragraph{Confusion Extremes}

Multiple type pairs achieved perfect separation (0\% error rate) in at least one model, with several pairs consistently separable across all classifiers, as shown in Table~\ref{tab:perfect-error-rate-zero}. Notably, \texttt{phishing vs. worm} and \texttt{iframe vs. worm} were perfectly classified by all five models, highlighting strong static differences between these types. Other pairs, such as \texttt{adware vs. injector}, achieved perfect separation only with specific classifiers like Random Forest. These results illustrate that while certain malware types are universally distinguishable, others require more advanced models to achieve optimal separation.

\begin{table}[h]
\centering
\small
\begin{tabular}{|c|c|}
\hline
\textbf{Type Pair} & \makecell{\textbf{Models with} \\ \textbf{0\,\% Error rate}} \\
\hline
autorun vs clicker                      & 5 \\
iframe vs injector                      & 5 \\
redirector vs worm                      & 5 \\
phishing vs worm                        & 5 \\
autorun vs phishing                     & 5 \\
autorun vs redirector                   & 5 \\
injector vs phishing                    & 5 \\
iframe vs worm                          & 5 \\
autorun vs iframe                       & 5 \\
injector vs redirector                  & 5 \\
backdoor vs redirector                  & 4 \\
clicker vs virus                        & 1 \\
adware vs injector                      & 1 \\
\hline
\end{tabular}
\caption{Number of classifiers achieving perfect separation (0\,\% error rate) for each malware type pair.}
\label{tab:perfect-error-rate-zero}
\end{table}

\paragraph{Summary}
The inter-type classification experiments showed that Random Forest and XGBoost consistently delivered the best results in distinguishing between malware types, achieving superior accuracy and reliable performance across varied type pairs. Random Forest reached an average accuracy of 97.5\% and moderate training time ($0.79\,\text{s}$). XGBoost followed closely with 97.4\% accuracy but required significantly longer training ($20.5\,\text{s}$), making it a more resource-intensive choice despite strong performance.

MLP delivered solid results (97.2\% accuracy) with a moderate training cost ($3.8\,\text{s}$), while KNN, although the fastest model ($0.04\,\text{s}$), showed high variability (standard deviation $0.07$) and the highest error rate (3.4\%), limiting its effectiveness in more complex distinctions.

Some type pairs, such as \texttt{autorun} vs. \texttt{worm} and \texttt{injector} vs. \texttt{downloader}, were particularly difficult to separate, with misclassification rates up to 44.5\%. Conversely, 13 pairs, including \texttt{phishing\hspace{-0.2em} vs.\hspace{-0.2em} worm} and \texttt{redirector\hspace{-0.2em} vs.\hspace{-0.2em} worm}, achieved perfect separability (0\% error rate) across all models, indicating strong static differences. These results confirm that while many types are distinguishable with high accuracy, certain categories present consistent challenges across models, emphasizing the importance of model choice and feature robustness.\\

\noindent\textbf{Classification Inter-Families}
\paragraph{Per-Model Performance Summary}
Across all inter-family classification tasks, Random Forest and XGBoost again proved to be the most effective classifiers. Random Forest achieved the highest average accuracy ($93.7\,\%$) and the lowest error rate ( $6.3\,\%$). It also demonstrated strong consistency (standard deviation of accuracy $0.072$) while maintaining a reasonable training time ($0.80\,\text{s}$).

XGBoost closely followed with an average accuracy of $93.4\,\%$ and error rate of $6.6\,\%$. Although it exhibited slightly higher variability ($ 0.076$), it benefited from a faster average training time of $0.27\,\text{s}$, making it a highly competitive alternative.

MLP and KNN showed comparable accuracy (both $92.5\,\%$), with KNN displaying greater variability ($0.080$), despite being the fastest model ($0.03\,\text{s}$). MLP was more stable but required significantly more training time ($1.97\,\text{s}$). SVM lagged behind with the lowest accuracy ($91.6\,\%$) and highest standard deviation ($0.095$), indicating less reliable classification across families.

\begin{table*}[h]
\centering
\begin{tabular}{|l|c|c|c|c|}
\hline
\textbf{Model} & \textbf{Avg Accuracy} & \textbf{Std Accuracy}  & \textbf{Avg Training Time (s)} \\
\hline
KNN           & 0.9250 & 0.0801  & 0.0278 \\
MLP           & 0.9246 & 0.0847  & 1.9703 \\
SVM           & 0.9163 & 0.0954   & 0.0932 \\
XGBoost       & 0.9342 & 0.0757   & 0.2686 \\
RandomForest  & 0.9370 & 0.0724   & 0.8042 \\
\hline
\end{tabular}%

\caption{Summary of average model performance in interclass family-based experiments.}
\label{tab:interclass-fam-summary}
\end{table*}

In Table~\ref{tab:interclass-fam-summary}, the average accuracy, and training time are computed over all pairs of malware families.

\paragraph{Family Separability Observations}

Some malware families were found to be particularly challenging to distinguish from each other. For instance, Agensla and Taskun showed substantial confusion, with over 40\% of Agensla samples misclassified as Taskun. Similarly, families like Androm and Agensla, as well as Noon and Agensla, exhibited high overlap, with misclassification rates around 40\%, as can be seen in Table \ref{tab:family-separability}.

In contrast, 5 family pairs were perfectly separable across all classifiers. Pairs such as Makoob vs. Taskun, Makoob vs. Mokes, and DCRat vs. Makoob achieved zero error rate, indicating strong static feature differences. Overall, these results suggest that while some families share structural similarities, others are highly distinct, allowing for reliable classification.

\begin{table}[h]
\centering
\begin{tabular}{|l|c|}
\hline
\textbf{Family Pair} & \textbf{Misclassification Rate} \\
\hline
Agensla $\rightarrow$ Taskun & 40.6\% \\
Androm $\rightarrow$ Agensla & 40.4\% \\
Noon $\rightarrow$ Agensla & 39.2\% \\
Remcos $\rightarrow$ Agensla & 38.1\% \\
Taskun $\rightarrow$ Noon & 35.2\% \\
\hline
Makoob $\leftrightarrow$ Taskun & 0.0\% \\
Makoob $\leftrightarrow$ Mokes & 0.0\% \\
Makoob $\leftrightarrow$ Seraph & 0.0\% \\
DCRat $\leftrightarrow$ Makoob & 0.0\% \\
Crysan $\leftrightarrow$ Makoob & 0.0\% \\

\hline
\end{tabular}
\caption{Misclassification rates for selected family pairs in interclass classification.}
\label{tab:family-separability}
\end{table}

\begin{figure}
        \includegraphics[width=\columnwidth]{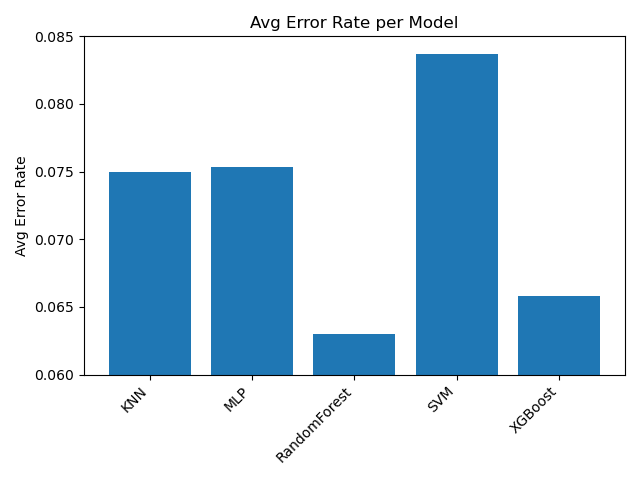}
        \caption{Average false error rate across all inter-family tasks.}
        \label{fig:interfam-fpr}
\end{figure}

\begin{figure}
        \includegraphics[width=\columnwidth]{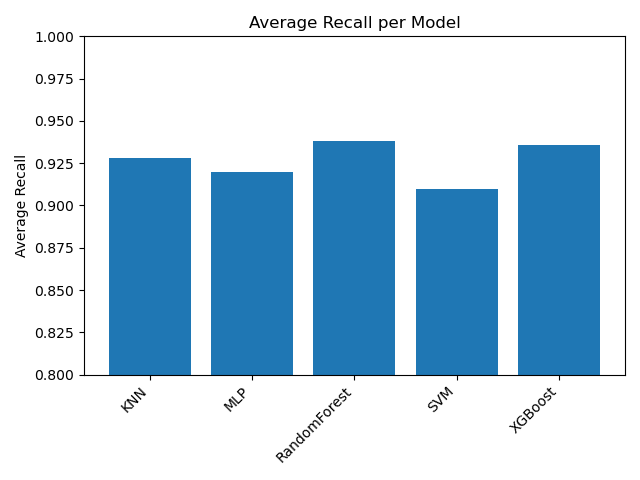}
        \caption{Average recall across all inter-family tasks.}
        \label{fig:interfam-recall}
    
\end{figure}

\begin{figure}
        \includegraphics[width=\columnwidth]{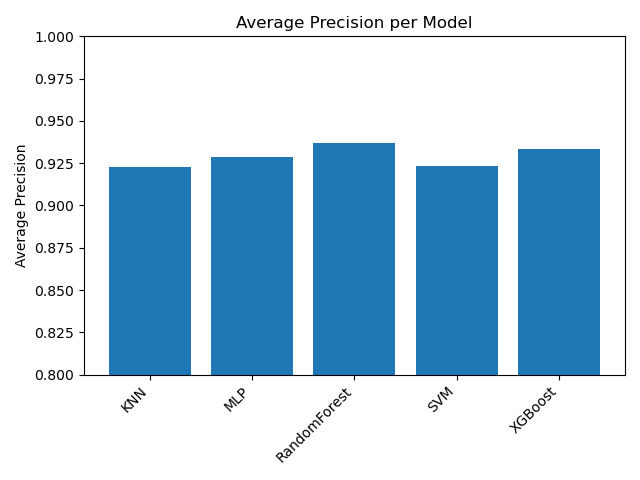}
        \caption{Average precision across all inter-family tasks.}
        \label{fig:interfam-precision}
\end{figure}

Figure~\ref{fig:interfam-fpr} presents the average error rates achieved by each model during inter-family classification. In addition, Figure~\ref{fig:interfam-precision} highlights model precision across family pairs, while Figure~\ref{fig:interfam-recall} illustrates recall performance. These visual summaries provide further insight into the relative strengths and weaknesses of each classifier when tasked with distinguishing between closely related malware families.

    

\paragraph{False Positive and Confusion Extremes}
The highest error rate was 39.43\% in the \texttt{Agensla vs.\ Androm} classification using SVM, indicating substantial confusion between these two families under the SVM decision boundary.
 This significant confusion suggests strong feature similarity between these two families, particularly under the SVM classifier, which struggled with their differentiation.

In contrast, numerous family pairs, such as \texttt{Crysan vs. Makoob}, \texttt{Makoob vs. Seraph}, and \texttt{Makoob vs. Mokes}, achieved perfect separability (0\% error rate) across all five models. These results underline the fact that while some malware families share common static characteristics, others remain distinct enough to allow classifiers to achieve flawless discrimination.

A selection of the most and least confused family pairs is summarized in Table~\ref{tab:zero-error-counts}, providing further insight into which family-level distinctions posed the greatest challenges or offered the clearest separability.

\begin{table}[h]
\centering
\small
\begin{tabular}{|l|c|}
\hline
\textbf{Type Pair} & \makecell{\textbf{Models with} \\ \textbf{0\,\% Error rate}} \\
\hline
Crysan vs Makoob                     & 5  \\ 
Makoob vs Mokes                      & 5  \\ 
Makoob vs Seraph                     & 5  \\ 
DCRat vs Makoob                      & 5  \\
Makoob vs Taskun                     & 5  \\
Taskun vs Zenpak                      & 4  \\ 
Makoob vs Strab                      & 3  \\ 

\hline
\end{tabular}
\caption{Number of classifiers achieving zero error rate for each family pair.}
\label{tab:zero-error-counts}
\end{table}

\paragraph{Summary} The inter-family classification experiments revealed that distinguishing between malware families is generally more challenging than between types. The average accuracy across all models was lower than in the inter-type experiments, with Random Forest and XGBoost achieving approximately 93.7\% and 93.4\% accuracy, respectively, compared to their performance exceeding 97\% in inter-type classification.

Furthermore, error rates were consistently higher in family-based tasks, suggesting greater overlap in static features between certain families, as shown in Table \ref{tab:fpr-type-vs-family}. 

\begin{table*}[h]
\centering
\begin{tabular}{|l|c|c|}
\hline
\textbf{Model} & \textbf{Avg Error Rate (Types)} & \textbf{Avg Error Rate (Families)} \\
\hline
KNN           & 0.033746 & 0.074974  \\
MLP           & 0.027799 & 0.075370 \\
SVM           & 0.029974 & 0.083660  \\
XGBoost       & 0.026134 & 0.065791 \\
RandomForest  & 0.024998  & 0.062969  \\
\hline
\end{tabular}
\caption{Comparison of average error rates between type-based and family-based interclass classification.}
\label{tab:fpr-type-vs-family}
\end{table*}

While several family pairs were still easily separable, particularly those involving \texttt{Makoob}, the overall trend indicates that family-level discrimination requires more nuanced modeling or enriched feature sets to achieve the same level of performance as type-based classification.

\subsection{Multiclass Classification Experiments}
While binary classification provides insights into pairwise or category-vs-benign detection scenarios, real-world malware detection systems often require the ability to classify samples into multiple categories simultaneously. Multiclass classification addresses this need by training models to predict the correct category label from a broader set of possible malware types or families.

This section presents experiments where machine learning models were trained and evaluated on multiclass datasets, aiming to assign each sample to its respective malware type or family. The focus is on assessing the overall classification accuracy, per-class precision and recall, confusion patterns between classes, and model robustness when exposed to more complex, multi-label decision boundaries. 

\subsubsection{Experimental Setup}
The multiclass experiments followed a similar evaluation pipeline to the binary classification tasks, with the following key differences:

\begin{itemize}
    \item \textbf{Dataset Composition.} 
    Each multiclass dataset was constructed by randomly selecting approximately 1,000 samples from each malware category. This uniform sampling ensured balanced class distributions, with approximately 14,000 samples for type-based classification and 17,000 samples for family-based classification. The datasets were then split into training and test sets using the same 66/33 ratio as in binary experiments. 

    \item \textbf{Label Encoding.} 
    Instead of binary labels \{0, 1\}, each sample was assigned a multiclass label from the set \{0, 1, ..., $n$--1\}, where $n$ is the total number of classes (14 for types, 17 for families).


    \item \textbf{Feature Selection.}
    During training, the feature selection process evaluated a broader set of $k$ values to accommodate multiclass complexity:
    \[
    k \in \{50, 100, 200, 300, 400, 500\}
    \]


    \item \textbf{Multiclass Classification Models:}
    All classifiers were configured to support multiclass:
    \begin{itemize}
        \item SVM used a one-vs-one (OvO) approach internally \cite{scikitlearnSVM}. 
        \item Random Forest, XGBoost, MLP, and KNN supported multiclass natively.
    \end{itemize}
\end{itemize}

The remaining elements of the pipeline, including feature standardization, hyperparameter tuning, and evaluation strategy, remained consistent with the binary classification setup.

\subsubsection{Type-based Multiclass Classification}
\label{sub: Type-based Multiclass Classification}

\paragraph{Per-Model Performance Summary}  
In the multiclass classification of malware types, SVM achieved the highest overall accuracy (81.1\%), outperforming other models in both macro-averaged precision (83.0\%) and F1-score (81.0\%). Despite this, it maintained a moderate training time ($14.3\,\text{s}$), making it an efficient and effective choice for multiclass tasks.

MLP followed closely with an accuracy of 80.1\%, offering strong recall (80.4\%) but at the cost of the longest training time among all models ($51.4\,\text{s}$). Random Forest delivered comparable results (accuracy 79.6\%, precision 81.3\%) with more balanced computational demands ($25.9\,\text{s}$).

KNN, although fast (training time $0.69\,\text{s}$), achieved lower accuracy\\(78.5\%) and the highest false positive rates across several classes, highlighting its limitations in handling multiclass distinctions. XGBoost, typically strong in binary tasks, ranked last in this setting with 78.2\% accuracy and relatively long training times ($48.3\,\text{s}$), suggesting that its advantage may diminish under multiclass complexity, as can be seen in Table~\ref{tab:multiclass-type-summary}.

\begin{table*}[h]
\centering
\resizebox{\textwidth}{!}{%
\begin{tabular}{|l|c|c|c|c|c|}
\hline
\textbf{Model} & \textbf{Accuracy} & \textbf{Macro Precision} & \textbf{Macro Recall} & \textbf{Macro F1-score} & \textbf{Train Time (s)} \\
\hline
KNN           & 0.7846 & 0.7975 & 0.7873 & 0.7832 & 0.6880 \\
MLP           & 0.8009 & 0.8154 & 0.8037 & 0.7988 & 51.4171 \\
RandomForest  & 0.7957 & 0.8133 & 0.7990 & 0.7949 & 25.9076 \\
SVM           & 0.8110 & 0.8305 & 0.8127 & 0.8104 & 14.2955 \\
XGBoost       & 0.7824 & 0.7842 & 0.7849 & 0.7839 & 48.2814 \\
\hline
\end{tabular}%
}
\caption{Summary of average model performance in type-based multiclass classification.}
\label{tab:multiclass-type-summary}
\end{table*}

\paragraph{Class-Level Observations}  
Classification performance varied notably\\ across malware types. The best-recognized classes were \texttt{phishing}, \texttt{coinminer}, and \texttt{clicker}, all achieving average recall above 90\%. Specifically, \texttt{phishing} reached a near-perfect recall of 99\% across all models, with minimal false positives and false negatives, confirming its distinct static features.

In contrast, classes like \texttt{worm}, \texttt{virus}, and \texttt{downloader} were significantly harder to classify. \texttt{Worm} achieved the lowest average recall (49\%), often being confused with \texttt{autorun}, while \texttt{virus} and \texttt{downloader} followed with recall rates of 51\% and 57\%, respectively. These lower-performing categories also exhibited higher false positive and false negative rates, indicating structural overlap with other classes.

\begin{table*}[h]
\centering
\resizebox{0.8\textwidth}{!}{%
\begin{tabular}{|l|c|c|c|c|c|}
\hline
\textbf{Class}    & \textbf{Avg Precision} & \textbf{Avg Recall} & \textbf{Avg F1-score} & \textbf{Avg FPR} & \textbf{Avg FNR} \\
\hline
phishing         & 0.996                  & 0.995               & 0.995                  & 0.004            & 0.005            \\
coinminer        & 0.978                  & 0.938               & 0.957                  & 0.022            & 0.062            \\
clicker          & 0.915                  & 0.914               & 0.915                  & 0.085            & 0.086            \\
spyware          & 0.812                  & 0.901               & 0.854                  & 0.188            & 0.099            \\
adware           & 0.809                  & 0.896               & 0.850                  & 0.191            & 0.104            \\
\hline
downloader       & 0.687                  & 0.567               & 0.618                  & 0.313            & 0.433            \\
injector         & 0.680                  & 0.748               & 0.710                  & 0.320            & 0.252            \\
autorun          & 0.511                  & 0.810               & 0.623                  & 0.489            & 0.190            \\
virus            & 0.802                  & 0.510               & 0.619                  & 0.198            & 0.490            \\
worm             & 0.702                  & 0.493               & 0.575                  & 0.298            & 0.507            \\
\hline
\end{tabular}%
}
\caption{Class-level performance metrics including average false positive rate (FPR) and false negative rate (FNR).}
\label{tab:multiclass-type-class-metrics}
\end{table*}

Each entry in Table \ref{tab:multiclass-type-class-metrics} is the simple arithmetic mean of the corresponding per-class metric across our five classifiers. In other words, for each malware class we first computed its precision, recall, F1-score, false positive rate (FPR) and false negative rate (FNR) on the held-out test set for each model, then averaged those five values to obtain the "Avg ..." figures reported here. Because each class was represented almost equally in our multiclass test splits, these unweighted averages give a clear picture of typical per-class performance across classifiers.

\paragraph{Confusion Analysis}
Detailed confusion analysis revealed consistent patterns of misclassification among certain malware types. The most confused pair was \texttt{worm} misclassified as \texttt{autorun}, with 694 instances, accounting for 44.1\% of all \texttt{worm} samples. Similarly, \texttt{downloader} was often mistaken for \texttt{injector} (578 cases, 32.9\% of \texttt{downloader}) and \texttt{virus} for \texttt{autorun} (545 cases, 29.7\% of \texttt{virus}).

On the other hand, minimal confusion was observed between some type pairs, with the lowest misclassification rates at only 0.06\%, as summarized in Table~\ref{tab:multiclass-type-confusion}. For example, \texttt{iframe} was misclassified as \texttt{phishing}, or \texttt{phishing} as \texttt{virus}, only once across all predictions.

\begin{table*}[h]
\centering
\resizebox{0.7\textwidth}{!}{%
\begin{tabular}{|l|c|c|}
\hline
\textbf{Confusion Pair} & \textbf{Misclassifications} & \textbf{Misclassification Rate (\%)} \\
\hline
\texttt{worm} $\rightarrow$ \texttt{autorun} & 694 & 44.1\% of worm \\
\texttt{downloader} $\rightarrow$ \texttt{injector} & 578 & 32.9\% of downloader \\
\texttt{virus} $\rightarrow$ \texttt{autorun} & 545 & 29.7\% of virus \\
\texttt{injector} $\rightarrow$ \texttt{downloader} & 404 & 24.6\% of injector \\
\texttt{redirector} $\rightarrow$ \texttt{iframe} & 204 & 12.8\% of redirector \\
\hline
\texttt{iframe} $\rightarrow$ \texttt{phishing} & 1 & 0.06\% of iframe \\
\texttt{phishing} $\rightarrow$ \texttt{virus} & 1 & 0.06\% of phishing \\
\texttt{spyware} $\rightarrow$ \texttt{iframe} & 1 & 0.06\% of spyware \\
\texttt{autorun} $\rightarrow$ \texttt{downloader} & 1 & 0.06\% of autorun \\
\hline
\end{tabular}%
}
\caption{Most and least frequent misclassification pairs with percentages, aggregated across all models (type-based multiclass classification).}
\label{tab:multiclass-type-confusion}
\end{table*}

These misclassification counts are aggregated across all models, reflecting the overall confusion trends rather than the result of any single classifier.

\paragraph{Summary}
The type-based multiclass classification experiments revealed that while some models achieved solid macro precision, with SVM reaching 83.0\%, followed by MLP at 81.5\% and Random Forest at 81.3\%, the overall level of misclassification remained high for certain classes. This indicates that despite good overall precision, specific malware types are frequently confused.

In particular, \texttt{worm}, \texttt{virus}, and \texttt{downloader} presented consistent classification difficulties, with frequent misclassifications such as \texttt{worm} being labeled as \texttt{autorun} in 44.1\% of cases and \texttt{downloader} as \texttt{injector} in 32.9\% of cases. These confusion rates are considerably higher than those observed in the family-based multiclass classification, indicating that malware types, although fewer in number, exhibit more overlap in static features.

Conversely, classes with distinctive traits, such as \texttt{phishing} and \texttt{coinminer}, were consistently well-identified across models, achieving macro recall values near or above 0.95. These findings emphasize that while macro precision provides a useful high-level evaluation, understanding class-specific confusion is essential for assessing model robustness in real-world detection scenarios.

\subsubsection{Family-Based Multiclass Classification}

\begin{table*}[h]
\centering
\resizebox{\textwidth}{!}{%
\begin{tabular}{|l|c|c|c|c|c|}
\hline
\textbf{Model} & \textbf{Accuracy} & \textbf{Macro Precision} & \textbf{Macro Recall} & \textbf{Macro F1-score} & \textbf{Train Time (s)} \\
\hline
KNN           & 0.6868 & 0.6966 & 0.6884 & 0.6897 & 1.0594 \\
MLP           & 0.6552 & 0.6635 & 0.6575 & 0.6582 & 81.3922 \\
RandomForest  & 0.7334 & 0.7460 & 0.7353 & 0.7337 & 20.6245 \\
SVM           & 0.5991 & 0.6185 & 0.6026 & 0.5939 & 56.3692 \\
XGBoost       & 0.7339 & 0.7371 & 0.7361 & 0.7341 & 45.7264 \\
\hline
\end{tabular}%
}
\caption{Summary of average model performance in family-based multiclass classification.}
\label{tab:multiclass-family-summary}
\end{table*}

\paragraph{Per-Model Performance Summary}
Multiclass classification of malware families, summarized in Table~\ref{tab:multiclass-family-summary}, revealed a wider range of model performance compared to type-based tasks. XGBoost and Random Forest emerged as the most accurate classifiers, both achieving an average accuracy of approximately $73.4\%$. XGBoost slightly outperformed Random Forest in terms of overall accuracy and macro-averaged metrics, while maintaining a moderate training time ($45.7\,\text{s}$). 

KNN followed with an accuracy of $68.7\%$, offering fast training ($1.06\,\text{s}$) while maintaining strong performance. In contrast, MLP and SVM lagged behind, with MLP reaching $65.5\%$ accuracy at a high computational cost ($81.4\,\text{s}$), and SVM performing the worst with an accuracy of $59.9\%$.

\paragraph{Class-Level Observations}  
Classification performance varied notably across malware families. The best-recognized families were \texttt{Makoob}, \texttt{DCRat}, and \texttt{Strab}, all achieving high average recall, with \texttt{Makoob} reaching a near-perfect recall of 99\% across all models. This highlights its distinctive static features and minimal confusion with other families.

In contrast, families like \texttt{Noon}, \texttt{Agensla}, and \texttt{Androm} were significantly harder to classify. \texttt{Noon} achieved the lowest average recall (35\%), often being confused with \texttt{Taskun} and \texttt{Agensla}, while \texttt{Agensla} and \texttt{Androm} followed with recall rates of 41\% and 51\%, respectively. 

\begin{table*}[h]
\centering
\resizebox{0.8\textwidth}{!}{%
\begin{tabular}{|l|c|c|c|c|c|}
\hline
\textbf{Family}    & \textbf{Avg Precision} & \textbf{Avg Recall} & \textbf{Avg F1-score} & \textbf{Avg FP} & \textbf{Avg FN} \\
\hline
Makoob             & 0.936                  & 0.987               & 0.961                 & 21.4            & 4.2             \\
DCRat              & 0.976                  & 0.965               & 0.970                 & 7.8             & 11.6            \\
Strab              & 0.803                  & 0.833               & 0.816                 & 63.2            & 51.4            \\
Zenpak             & 0.742                  & 0.820               & 0.778                 & 91.2            & 57.2            \\
Seraph             & 0.618                  & 0.810               & 0.695                 & 180.2           & 66.4            \\
\hline
Noon               & 0.432                  & 0.346               & 0.382                 & 143.0           & 217.0           \\
Agensla            & 0.379                  & 0.410               & 0.391                 & 235.8           & 198.8           \\
Androm             & 0.614                  & 0.511               & 0.548                 & 99.6            & 157.6           \\
Taskun             & 0.477                  & 0.556               & 0.498                 & 237.2           & 144.4           \\
Remcos             & 0.776                  & 0.556               & 0.641                 & 60.8            & 156.8           \\
\hline
\end{tabular}%
}
\caption{Family-level performance metrics including average false positives and false negatives across all models.}
\label{tab:multiclass-family-class-metrics}
\end{table*}

Each entry in Table \ref{tab:multiclass-family-class-metrics} is the simple arithmetic mean of the corresponding per-class metric across our five classifiers. In other words, for each malware class we first computed its precision, recall, F1-score, false positive rate (FPR) and false negative rate (FNR) on the held-out test set for each model, then averaged those five values to obtain the "Avg ..." figures reported here. Since each class was represented almost equally in our multiclass test splits, these unweighted averages give a clear picture of typical per-class performance across classifiers.

\paragraph{Confusion Analysis}
The confusion analysis highlights specific family pairs that were frequently misclassified. The most severe confusion was observed between \texttt{Taskun} and \texttt{Agensla}, where $19.1\%$ of \texttt{Taskun} samples were misclassified as \texttt{Agensla}, and $18.4\%$ of \texttt{Agensla} as \texttt{Taskun}. Additionally, \texttt{Noon} exhibited substantial confusion, being misclassified as \texttt{Taskun} and \texttt{Agensla} in $18.1\%$ and $15.2\%$ of cases, respectively.

Other notable confusion pairs included \texttt{Mokes} misclassified as  \texttt{Zenpak}~($12.8\%$),~and \texttt{Injuke} misclassified as \texttt{Seraph} ($8.6\%$) and \texttt{Convagent} ($8.2\%$). These overlaps suggest that some families share common static characteristics, complicating accurate identification.

In contrast, several family pairs were rarely confused. For example, \texttt{Crypt} $\rightarrow$ \texttt{Zenpak}, \mbox{\texttt{Strab} $\rightarrow$ \texttt{Crysan}}, and \texttt{Remcos} $\rightarrow$ \texttt{DCRat} each showed only a~single misclassification, representing less than $0.1\%$ of the respective true class. The most and least frequent misclassification pairs are summarized in Table \ref{tab:multiclass-family-confusion}.

\begin{table*}[h]
\centering
\resizebox{0.7\textwidth}{!}{%
\begin{tabular}{|l|c|c|}
\hline
\textbf{Confusion Pair} & \textbf{Misclassifications} & \textbf{Misclassifications Rate (\%)} \\
\hline
\texttt{Taskun} $\rightarrow$ \texttt{Agensla} & 310 & 19.1\% of Taskun \\
\texttt{Agensla} $\rightarrow$ \texttt{Taskun} & 310 & 18.4\% of Agensla \\
\texttt{Noon} $\rightarrow$ \texttt{Taskun}    & 300 & 18.1\% of Noon \\
\texttt{Noon} $\rightarrow$ \texttt{Agensla}   & 252 & 15.2\% of Noon \\
\texttt{Mokes} $\rightarrow$ \texttt{Zenpak}   & 218 & 12.8\% of Mokes \\
\hline
\texttt{Crypt} $\rightarrow$ \texttt{Zenpak}   & 1 & 0.06\% of Crypt \\
\texttt{Strab} $\rightarrow$ \texttt{Crysan}   & 1 & 0.07\% of Strab \\
\texttt{Remcos} $\rightarrow$ \texttt{DCRat}   & 1 & 0.06\% of Remcos \\
\hline
\end{tabular}%
}
\caption{Most and least frequent misclassification pairs with percentages, aggregated across all models (family-based multiclass classification).}
\label{tab:multiclass-family-confusion}
\end{table*}

\paragraph{Summary}
The family-based multiclass classification experiments showed that, although models achieved lower macro precision compared to type-based classification (e.g., best macro precision $0.75$ for Random Forest vs.\ $0.83$ for SVM in type-based), the most extreme misclassification rates between individual families were notably lower.

This implies that, despite the greater challenge in achieving high overall precision when classifying malware families, the most problematic family pairs were less frequently confused compared to the most problematic type pairs. 

These findings suggest that families, while harder to classify with high precision across the board, often exhibit clearer boundaries in the most extreme confusion cases. Future improvements in model precision could thus yield stronger gains in family-based classification.

\section{Conclusions}

This work set out to address one of the central challenges in machine-learning-based malware detection: the lack of publicly available datasets that provide not only feature vectors and binary samples, but also include detailed type and family annotations. To this end, a comprehensive malware dataset was constructed by combining multiple sources. Precomputed type labels were obtained from ClarAVy for binaries hosted on VirusShare, while malware families were collected from VX Underground, with family labels parsed directly from filenames. In total, the final dataset covered 14 malware types and 17 malware families, providing both feature-level and binary-level access to the samples. The resulting dataset is released publicly for research purposes on the \url{https://github.com/CS-and-AI/RawMal-TF}.

A uniform feature extraction pipeline was developed using the EMBER framework, which enabled the consistent transformation of binaries into numerical representations. The feature pipeline included customized preprocessing steps to align raw features with the EMBER schema, as well as integration of benign samples from the original EMBER dataset to support malware vs. benign experiments.

The experimental evaluation was divided into several parts. Binary classification experiments were conducted to distinguish between malware and benign files, both at the type and family levels. In full-scope settings, Random Forest and XGBoost achieved near-perfect performance, with average accuracies exceeding 98.5\% on both type-based and family-based datasets. Truncated datasets, where each category was limited to approximately 1,000 malware and benign samples, showed only a minor drop in performance, confirming the robustness of static features even under reduced data conditions.

In interclass classification experiments, the models were tasked with discriminating between pairs of malware types or families. While type-based classification reached average accuracies around 97.5\% (Random Forest) and 97.4\% (XGBoost), family-based classification proved more challenging, with average accuracies dropping to about 93.7\% and 93.4\%, respectively. 

In multiclass experiments, while type-based classification yielded higher overall accuracy (SVM at 81.1\%) compared to family-based classification (XGBoost and Random Forest at around 73.4\%), an important observation emerged: despite lower overall precision in the family-based case, individual family pairs were less frequently confused than type pairs. 


The dataset and experiments described in this work provide a foundation for further research into improving classification performance, optimizing feature extraction, and exploring additional approaches such as dynamic analysis or hybrid detection strategies.

\backmatter

%

\section*{Declarations}

The authors have no relevant financial or non-financial interests to disclose.

\bibliography{sn-article}


\end{document}